\documentclass[sigplan]{acmart}

\usepackage{booktabs}   
\usepackage{subcaption} 

\usepackage{amsmath}    
\usepackage{caption}
\usepackage{listings}
\usepackage{xcolor}
\usepackage{graphicx}
\usepackage{xspace}
\usepackage{algorithm}
\usepackage[noend]{algpseudocode}
\usepackage{multirow}
\usepackage[export]{adjustbox}
\usepackage{enumitem}
\usepackage{float}

\newfloat{algorithm}{t}{lop}

\algnewcommand{\IIf}[1]{\State\algorithmicif\ #1\ \algorithmicthen}
\algnewcommand{\EndIIf}{\unskip\ \algorithmicend\ \algorithmicif}

\newcommand{\papertitle}

\newcommand{\sparse}[1]{{\textbf{\textit{#1}}}}
\newcommand{\dense}[1]{{\textbf{#1}}}

\newcommand{\etal}{\textit{et al.}}

\newcommand{\ie}{\textit{i.e.,}}

\newcommand{\rulesep}{\unskip\ \vrule\ }
\newcommand{\WRP}{\par\qquad\(\hookrightarrow\)\enspace}

\graphicspath{ {./images/} }

\newcommand{\system}{SparseLNR\xspace}
\newcommand{\code}[1]{\texttt{#1}} 

\copyrightyear{2022}
\acmYear{2022}
\setcopyright{rightsretained}
\acmConference[ICS '22]{2022 International Conference on Supercomputing}{June 28--30, 2022}{Virtual Event, USA}
\acmBooktitle{2022 International Conference on Supercomputing (ICS '22), June 28--30, 2022, Virtual Event, USA}
\acmDOI{10.1145/3524059.3532386}
\acmISBN{978-1-4503-9281-5/22/06}

\begin{document}

\title[SparseLNR]{SparseLNR: Accelerating Sparse Tensor Computations Using Loop Nest Restructuring}         


\author{Adhitha Dias}
\orcid{0000-0003-3500-7547}             
\affiliation{
  \department{Electrical and Computer Engineering}              
  \institution{Purdue University}            
  \streetaddress{610 Purdue Mall}
  \city{West Lafayette}
  \state{IN}
  \postcode{47906}
  \country{USA}                    
}
\email{kadhitha@purdue.edu}          

\author{Kirshanthan Sundararajah}
\affiliation{
  \department{Electrical and Computer Engineering}              
  \institution{Purdue University}            
  \streetaddress{610 Purdue Mall}
  \city{West Lafayette}
  \state{IN}
  \postcode{47906}
  \country{USA}                    
}
\email{ksundar@purdue.edu}          

\author{Charitha Saumya}
\affiliation{
  \department{Electrical and Computer Engineering}              
  \institution{Purdue University}            
  \streetaddress{610 Purdue Mall}
  \city{West Lafayette}
  \state{IN}
  \postcode{47906}
  \country{USA}                    
}
\email{cgusthin@purdue.edu}          

\author{Milind Kulkarni}
\affiliation{
  \department{Electrical and Computer Engineering}              
  \institution{Purdue University}            
  \streetaddress{610 Purdue Mall}
  \city{West Lafayette}
  \state{IN}
  \postcode{47906}
  \country{USA}                    
}
\email{milind@purdue.edu}          


\begin{abstract}
Sparse tensor algebra computations have become important in many real-world applications like machine learning, scientific simulations, and data mining. 
Hence, automated code generation and performance optimizations for tensor algebra kernels are paramount. 
Recent advancements such as the Tensor Algebra Compiler (TACO) greatly generalize and automate the code generation for tensor algebra expressions.
However, the code generated by TACO for many important tensor computations remains suboptimal due to the absence of a scheduling directive to support transformations such as {\em distribution/fusion}.

This paper extends TACO's scheduling space to support kernel distribution/loop fusion in order to reduce asymptotic time complexity and improve locality of complex tensor algebra computations. 
We develop an intermediate representation (IR) for tensor operations called {\em branched iteration graph} which specifies breakdown of the computation into smaller ones (kernel distribution) and then fuse (loop fusion) outermost dimensions of the loop nests, while the innermost dimensions are distributed, to increase data locality.
We describe exchanges of intermediate results between space iteration spaces, transformation in the IR, and its programmatic invocation.
Finally, we show that the transformation can be used to optimize sparse tensor kernels. 
Our results show that this new transformation significantly improves the performance of several real-world tensor algebra computations compared to TACO-generated code.
\end{abstract}



\maketitle

\section{Introduction} \label{introduction}

Sparse tensor algebra is used in many machine learning domains such as
graph neural networks~\cite{fusedMM,FeatGraph,hamilton2018inductive}. 
Tensors are a generalization of matrices and are typically represented using n-dimensional arrays. 
However, when used to represent large graph-like structures, representing a tensor with a dense array is wasteful, as most values in the tensor are zero. In such cases, programmers use {\em compressed} representations of these {\em sparse tensors}.

The problem of compiler optimizations for sparse codes is well known 
~\cite{kjolstad:2017:taco,aartbik:93,michelle2015,compiler_in_mlir,pingali97}, and there are several challenges that compilers face: (1) tensor computations have to deal with specific data formats; (2) load imbalance can arise due to irregular structure; and 
(3) data locality issues arise due to the sparsity of the tensors.
TACO provides a compiler for automatically generating kernels for dense and sparse tensor algebra 
operations~\cite{kjolstad:2017:taco}. The tensor application is expressed in terms of three languages: a tensor algebra language for expressing the computation
(Section~\ref{tensor-index-notation}), a data representation language for specifying how sparse tensors are compressed, and a
scheduling language that specifies the schedule of the computation (Section~\ref{scheduling_primitives}).

The scheduling language provides the ability to define different schedules for computations depending on tensors' dimensionality and sparsity patterns, because one schedule may not fit all data formats and datasets. 
This allows the separation of {\em algorithmic} specification from the {\em scheduling} details of the computation.
Once both are specified, code can be generated to implement the desired algorithm and schedule.

One important consequence of TACO's code generation is that the asymptotic complexity of the kernels grows with the number of index variables in the tensor index notation~\cite{peterahrens}. 
For example, the complexity of ${\sparse A_{ij}} = \sum\nolimits_{k} {\sparse B_{ij}}\cdot C_{ik}\cdot D_{jk}$\footnote{Highlighted tensors denote sparse tensors.} is $O(nnz(B_{IJ})K)$\footnote{$nnz(B_{IJ})$ denotes the nonzero values of the sparse tensor B bounded by the hierarchical accesses $i$ and $j$.}, where $B$ is sparse.
If this example is extended with an additional computation, as in $A_{il} =\sum\nolimits_{kj} {\sparse B_{ij}}\cdot C_{ik}\cdot D_{jk}\cdot E_{jl}$, then the complexity is $O(nnz(B_{IJ})KL)$~\footnote{$K$ and $L$ denote the number of iterations or the dimensionality of $k$ and $l$ dimensions respectively.}---and this complexity increases with each additional index variable.
Hence, with increasing terms in the tensor expression, the asymptotic complexity of the resulting code blows up.

Interestingly, this asymptotic blowup is a consequence of doing multiple tensor operations in a single kernel. 
The computation could instead be expressed as two separate kernels, with the result of the first computation stored in a temporary tensor: ${\sparse T_{ij}} =\sum\nolimits_{k} {\sparse B_{ij}}\cdot C_{ik}\cdot D_{jk} $; $A_{il} =\sum\nolimits_{j} {\sparse T_{ij}}\cdot E_{jl}$. 
This computation has a complexity of $O(nnz(B_{IJ})(K+L))$. However, writing complex computations as {\em separate} TACO expressions has two downsides. 
First, it is no longer possible to apply schedule transformations, such as outer-loop parallelization, across the entire computation. 
Second, if the computations require large temporaries, materializing them results in performance degradation due to exhaustion of the last-level cache.

The correct schedule looks like neither the single-kernel approach nor the separate-kernels approach. 
Instead, it performs a single outer loop over the $i$ and $j$ indices and then performs the inner loop of the first kernel, stores the results in a temporary, then uses those results in the inner loop of the second kernel. 
This approach has an asymptotic complexity of $O(nnz(B_{IJ})(K+L))$, comparable to the separate kernel approach, but because the temporary is only live within the inner loops, it is much smaller and hence can fit in cache. 
Moreover, the overall computation is a single loop nest, allowing for the outer loops to be parallelized, tiled, etc.

The above schedule transformation is analogous to ones in {\em dense} tensor contraction that combine loop distribution and fusion to create imperfectly-nested loops~\cite{saday1}. 
But it is less clear how to use this technique on sparse loops for several reasons: 
(i) analysis is harder, because of the sparse tensor accesses and non-affine bounds, as polyhedral techniques do not work due to the use of dynamic array bounds in loops;
(ii) producing good schedules is harder because performance can degrade by forcing a sparse tensor to be processed using dense iteration; and
(iii) code generation is harder, as you need to deal with storage format-specific iteration machinery. 
For example, a sparse matrix and dense matrix multiplication (SpMM) may be performed with a sparse matrix of Compressed Sparse Row format (CSR), Coordinate format (COO), etc.~\cite{chau2018}. 
Hence, the compiler needs to tackle format-specific access patterns to generate code for SpMM for different storage formats. 

Our insight for tackling the complex scheduling transformations needed to avoid asymptotic blowup while preserving locality, is to use dense temporaries and introduce \textit{Sparse Loop Nest Restructuring (SparseLNR)\footnote{\url{https://github.com/adhithadias/SparseLNR}}} for tensor computations.
Crucially, these transformations can co-exist with TACO's other scheduling primitives~\cite{senanayake:2020:scheduling}. 

This paper introduces a new representation called {\em branched iteration graphs} that support imperfect nesting of sparse iteration. 
Given this representation, our compiler can restructure sparse tensor computations to remove the asymptotic blowup in sparse tensor algebra code generation while delivering good locality.
Our specific contributions are;

\begin{description}
\item[\textbf{Branched iteration graph for tensor multiplications}] We generalize the 
iteration graph intermediate representation (IR) of TACO to support imperfectly nested loop structures.

\item[\textbf{Branch IR transformation}] We design a sparse tensor transformation 
that transforms iteration graphs to express fusion and distribution.

\item[\textbf{New scheduling primitives}] We introduce a new scheduling primitive that lets programmers integrate fusion and distribution into TACO schedules.
\end{description}

For several real-world tensor algebra computations (Described in Section~\ref{benchmarks}) on various datasets (Shown in Table~\ref{tab:datasets}), using our new representation and transformations, we show that \system can achieve 1.23--1997x (single-thread) and 0.86--1263x (multi-thread) speedup over baseline TACO schedules, and 0.27--3.21x (single-thread) and 0.51--3.16x (multi-thread) speedup over TACO schedules of manually separated computations. 

\section{Background} 
\label{background}
This section provides the necessary background to understand sparse tensor algebra computations and different ways to schedule those computations.

\subsection{Tensor Index Notation} 
\label{tensor-index-notation}
Tensor index notation is a high-level representation used for describing tensor algebra expressions~\cite{kjolstad:2017:taco}.
Throughout the paper we will be using both the {\em standard
notation} and {\em tensor index notation} to denote tensor operations. 
For instance, the tensor computation $A_{ik} = \sum\nolimits_{j} {\sparse B_{ij}} C_{jk}$ written in standard notation is equivalent to $A(i,k) = {\sparse B(i,j)} * C(j,k)$, written in tensor index notation.\footnote{This computation is classic matrix-matrix multiply.} 
Here, all the tensors are matrices and indices $i,j,$ and $k$ are used to iterate over matrices $A, B,$ and $C$. 
In this computation, index $j$ must be iterated over the intersection of second dimension coordinates of $B$ and first dimension coordinates of $C$, whereas index $i$ and $k$ must be iterated over the first and second dimension coordinates of $B$ and $C$ respectively.

\subsection{Iteration Graph} 
\label{iteration_graph}

We first summarize TACO's iteration graph representation, which Kjolstad~\etal~describes in great detail~\cite{kjolstad:2017:taco}. 
When computing the tensor expression $ {\sparse A_{ij}} = \sum\nolimits_{k} {\sparse B_{ij}} C_{ik} D_{jk} $, coordinates $(i,j)$ of B, coordinates $(i,k)$ of C, and $(j,k)$ of D need to be iterated. 
An iterator on indices $(i,j,k)$ can iterate through all the coordinates of $B$, $C$, and $D$ and store the results in $A$. 
TACO represents the iteration space of a tensor expression using an iteration graph, an intermediate representation that defines tensor access patterns of indices.

\begin{figure}[!t]
    \vspace{-1em}
    \centering
    \hfill
    \begin{subfigure}[t]{0.3\columnwidth}
        \centering
        \includegraphics[scale=.3]{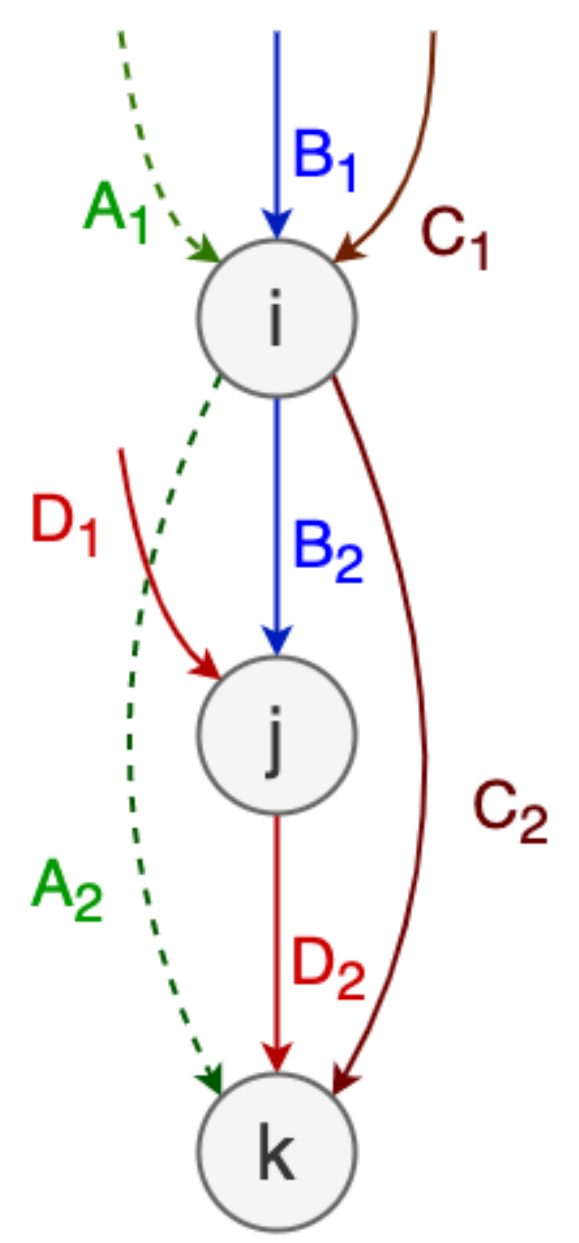}
        \caption{\ }
        \vspace{-10pt}
        \label{fig:sddmm-iter-graph}
        \Description[SDDMM original iteration graph]{<long description>}
    \end{subfigure}%
    \hfill
    \begin{subfigure}[t]{0.3\columnwidth}
        \centering
        \includegraphics[scale=.3]{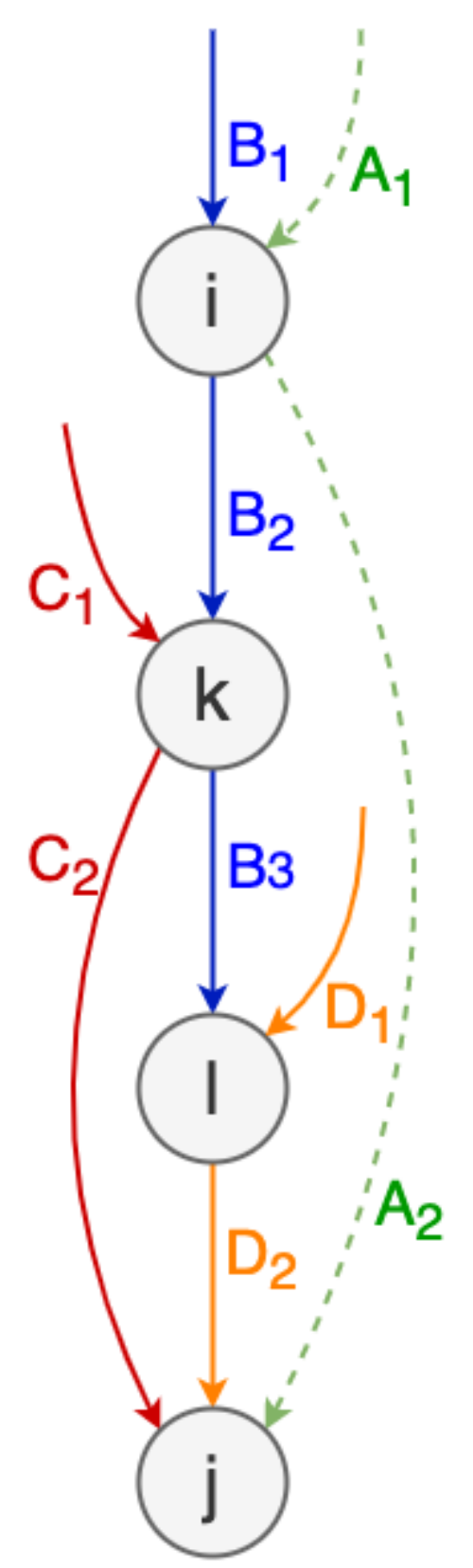}
        \caption{\ }
        \vspace{-10pt}
        \label{fig:mttkrp-iter-graph}
        \Description[MTTKRP original iteration graph]{<long description>}
    \end{subfigure}%
    \hfill
    \begin{subfigure}[t]{0.3\columnwidth}
        \centering
        \includegraphics[scale=.3]{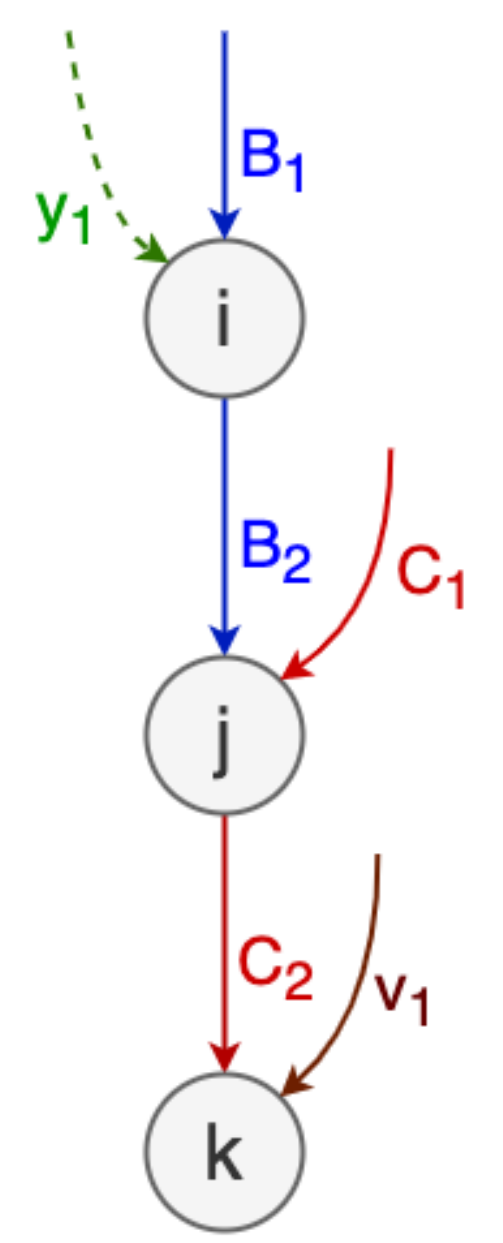}
        \caption{\ }
        \vspace{-10pt}
        \label{fig:spmv-spmv-iter-graph}
        \Description[sddmm-spmm original iteration graph]{<long description>}
    \end{subfigure}%
    \hfill
    \caption{Iteration graphs (a) SDDMM kernel $ {\sparse A_{ij}} = \sum\nolimits_{k} {\sparse B_{ij}} C_{ik} D_{jk} $ 
        (b) Khatri-Rao product (MTTKRP) kernel $ {\sparse A_{ik}} = \sum\nolimits_{kl} {\sparse B_{ikl}} C_{lj} D_{kj} $ 
        (c) Sparse matrix vector multiplication (SpMV) kernel preceded by another SpMV kernel $ y_{i} = \sum\nolimits_{jk} {\sparse B_{ij}} ({\sparse C_{jk}} v_{k}) $ }
    \label{fig:iteration-graphs}
    \vspace{-1.5em}
\end{figure}

Figure~\ref{fig:iteration-graphs} shows a few examples of iteration graphs. For example, a tensor expression $ {\sparse A_{ij}} = \sum\nolimits_{k} {\sparse B_{ij}} C_{ik} D_{jk} $ results in an iteration graph as shown in Figure~\ref{fig:sddmm-iter-graph} such that the indices lay in $i, j, k$ order.
Here, the order of $j$ and $k$ is not strict if C and D are dense. 
Figures~\ref{fig:mttkrp-iter-graph}~and~\ref{fig:spmv-spmv-iter-graph} are the iteration graphs of 
tensor expressions $ A_{ik} = \sum\nolimits_{kl} {\sparse B_{ikl}} C_{lj} D_{kj} $ and 
$ y_{i} = \sum\nolimits_{jk} {\sparse B_{ij}} {\sparse C_{jk}} v_{k} $ 
respectively.

Nodes in the iteration graph represent indices of tensor index notation. 
In other words, the iteration graph is a directed graph of these indices. 
These indices of the graph are topologically sorted such that it imposes sparse iteration constraints (\ie constraints that define the sparse tensor access patterns of indices due to lack of random access in general). 
Each index in the iteration graph can be expressed as a loop to iterate through a tensor. 
Therefore, a given tensor multiplication can be computed using nested loops, where each loop corresponds to an index variable in the iteration graph. 


\begin{definition} \label{iteration_graph_definition}
An iteration graph is a directed graph $ G = (V,P) $ where $ V = {v_1, v_2, ..., v_n} $ defines the set of index variables in the tensor index notation, and $ P = {p_1, p_2, ..., p_n} $ defines the set of tensor paths, a tensor path is a tuple of index variables associated with a particular tensor variable.
\end{definition} 
    

\subsection{Scheduling Primitives}\label{scheduling_primitives}

A tensor expression can have multiple valid schedules of computation as there are different valid orders of iterating through indices and multiple parallelization strategies. 
Kjolstad \etal~\cite{kjolstad:2017:taco} and 
Senanayake \etal~\cite{senanayake:2020:scheduling} have introduced scheduling primitives for tensor computations, with which the user can describe schedules to execute a given tensor computation. 
The scheduling primitives in TACO are the \textit{split} directive to split a loop into two loops for tiling, \textit{collapse} directive to collapse doubly nested loops into a single loop for balancing load among threads, \textit{reorder} directive\footnote{Also referred to as \textit{permute} directive in the literature.} to reorder loops, \textit{unroll} directive to perform loop unrolling, \textit{parallelize} directive to parallelize loops with OpenMP-based multithreaded execution (for outer loops) or vectorized execution (for inner loops).
Furthermore, Kjolstad \etal~\cite{kjolstad:2018:workspaces} 
added \textit{precompute} scheduling directive to use intermediate dense workspaces to remove sparse accesses when storing data values to output tensors.
\section{Overview} \label{overview}

There are a number of factors taken into account when deciding whether to apply transformations across kernels. 
If the working sets are small, running the kernels separately with good schedules defined on each individual kernel maybe faster than a fused kernel. 
But if the working sets are large resulting in large temporaries that do not fit in caches, it is better to fuse two kernels and try to maximize the data reuse by using the results produced by the first kernel and execute part of the second kernel without waiting for the completion of the first kernel. 



\begin{figure*}[t]
  \begin{adjustbox}{minipage=\linewidth,scale=0.9}
    \begin{subfigure}[t]{0.50\textwidth}
      \begin{lstlisting}[basicstyle=\small, gobble=8, tabsize=2, showtabs=false, showstringspaces=false]
        int32_t jY = 0;
        for (int32_t i = 0; i < C1_dimension; i++) {
          for (int32_t jB = B2_pos[i]; jB < B2_pos[(i + 1)]; jB++) {
            int32_t j = B2_crd[jB];
            double tkY_val = 0.0;
            for (int32_t k = 0; k < D2_dimension; k++) {
              tkY_val += B_vals[jB] * C_vals[i,k] * D_vals[j,k];
            }
            Y_vals[jY] = tkY_val;
            jY++;
          }
        }
      \end{lstlisting}
      \vspace{-10pt}
      \caption[For the list of figures]{$ {\sparse Y_{ij}} = \sum\nolimits_{k} {\sparse B_{ij}} C_{jk} D_{kj} $}
      \vspace{-10pt}
      \label{fig:sddmm-kernel}
      \Description[C++ code of the kernel sampled dense-dense matrix multiplication kernel]{<long description>}
    \end{subfigure}%
    \hfill%
    \hspace{2em}%
    \begin{subfigure}[t]{0.50\textwidth}
      \begin{lstlisting}[basicstyle=\small, gobble=8, tabsize=2, showstringspaces=false]
        for (int32_t i = 0; i < Y1_dimension; i++) {
          for (int32_t jY = Y2_pos[i]; jY < Y2_pos[(i + 1)]; jY++) {
            int32_t j = Y2_crd[jY];
            for (int32_t l = 0; l < E2_dimension; l++) {
              A_vals[i,l] = A_vals[i,l] + Y_vals[jY] * E_vals[j,l];
            }
          }
        }
      \end{lstlisting}
      \vspace{-10pt}
      \caption[For the list of figures]{$ A_{il} = \sum\nolimits_{j} {\sparse Y_{ij}} E_{jl} $}
      \vspace{-10pt}
      \label{fig:spmm-kernel}
      \Description[C++ code of the sparse matrix multiplication kernel]{<long description>}
    \end{subfigure}%
    \vskip\baselineskip
    \begin{subfigure}[t]{0.50\textwidth}
      \begin{lstlisting}[basicstyle=\small, gobble=8, tabsize=2, showstringspaces=false]
        for (int32_t i = 0; i < C1_dimension; i++) {
          for (int32_t jB = B2_pos[i]; jB < B2_pos[(i + 1)]; jB++) {
            int32_t j = B2_crd[jB];
            for (int32_t l = 0; l < E2_dimension; l++) {
              double tkA = 0.0;
              for (int32_t k = 0; k < D2_dimension; k++) {
                tkA += B_vals[jB]* C_vals[i,k]* D_vals[j,k]* E_vals[j,l];
              }
              A_vals[i,l] = A_vals[i,l] + tkA;
            }
          }
        }
      \end{lstlisting}
      \vspace{-10pt}
      \caption[For the list of figures]{$ A_{il} = \sum\nolimits_{jk} {\sparse B_{ij}} C_{jk} D_{kj} E_{jl} $}
      \vspace{-5pt}
      \label{fig:taco-fused-sddmm-spmm-kernel}
      \Description[C++ code of the kernel fused sampled dense-dense matrix multiplication kernel and sparse matrix multiplication kernel]{<long description>}
    \end{subfigure}%
    \hfill%
    \hspace{2em}%
    \begin{subfigure}[t]{0.50\textwidth}
      \begin{lstlisting}[basicstyle=\small, gobble=8, tabsize=2, showstringspaces=false]
        for (int32_t i = 0; i < C1_dimension; i++) {
          for (int32_t jB = B2_pos[i]; jB < B2_pos[(i + 1)]; jB++) {
            int32_t j = B2_crd[jB];
            double Y_val = 0.0;
            for (int32_t k = 0; k < D2_dimension; k++) {
              Y_val += B_vals[jB] * C_vals[i,k] * D_vals[j,k];
            }
            for (int32_t l = 0; l < E2_dimension; l++) {
              A_vals[i,l] = A_vals[i,l] + Y_val * E_vals[j,l];
            }
          }
        }
      \end{lstlisting}
      \vspace{-10pt}
      \caption[For the list of figures]{$ A_{il} = \sum\nolimits_{j} ( \sum\nolimits_{k} {\sparse B_{ij}} C_{jk} D_{kj} ) E_{jl}$}
      \vspace{-5pt}
      \label{fig:SparFF-fused-kernel}
      \Description[C++ code of the kernel fused sampled dense-dense matrix multiplication kernel and sparse matrix multiplication kernel]{<long description>}
    \end{subfigure}
  \end{adjustbox}
    \caption{Different schedules of executing $ A_{il} = \sum\nolimits {\sparse B_{ij}} \cdot C_{jk} \cdot D_{kj} \cdot  E_{jl} $. The code snippet~\ref{fig:spmm-kernel} executed immediately after the code snippet~\ref{fig:sddmm-kernel} computes the same result as fused operations explained in the code snippets~\ref{fig:taco-fused-sddmm-spmm-kernel} and~\ref{fig:SparFF-fused-kernel}. Here, the code snippet~\ref{fig:taco-fused-sddmm-spmm-kernel} has a perfectly nested loop structure while the code~\ref{fig:SparFF-fused-kernel} describes a nested loop structure for the same computation.}
    \vspace{-10pt}
    \label{fig:three graphs}
\end{figure*}

\subsection{Motivating Example} \label{motivating_example}

Consider the computation, $A = Sparse\ B \odot (CD) \cdot E$ that is used in graph embedding and graph neural networks~\cite{fusedMM,graph_embedding_networks}. 
The Hadamard product, or element-wise product, is denoted by $\odot$ and matrix multiplication is denoted by $\cdot$.
We can perform the above computation in the following order 
with fine-grained smaller tensor operations. 
$T = gemm (C, D)$, $Sparse\ U = spelmm (Sparse\ B, T)$, 
$A = spmm (Sparse\ U, E)$.
Here, $gemm$ stands for the generalized matrix multiplication, $spelmm$ stands for sparse element-wise multiplication, and $spmm$ stands for sparse matrix multiplication.
Materialization of these intermediate tensors leads to 
multiple issues:

\begin{enumerate}[align=left,leftmargin=*]
    \item Dense matrix multiplication results in 
    redundant calculations and unnecessary increase in asymptotic complexity, because later it is sampled by the Sparse B matrix.

    \item Values are produced long before they are consumed, which may cause them to be evicted from caches. 
	
    \item Having intermediate tensors is justifiable if intermediate results are needed for some other computation, nevertheless a single kernel maybe needed for faster operation.
\end{enumerate}

Introducing \textit{kernel fusion} to tensor computations can reduce these issues~\cite{fusedMM}.
In this section, we discuss different schedules for performing the computation $ A = Sparse\ B \cdot (CD) * E $, and motivate the need for supporting loop fusion for sparse tensor computations.

First, we discuss the opportunities for distribution in the running example using a fused kernel with high asymptotic complexity (Section~\ref{taco_original_kernel}).
Next, we discuss opportunities for fusion when the computation is split into two smaller kernels (Section~\ref{separate-kernel-execution-code}). 
Finally, in Section~\ref{subsection:fused-kernel-code} we discuss how we can exploit these different scenarios to construct a distributed (versus the fused kernel in Section~\ref{taco_original_kernel}) and then fused (as compared to the kernel in Section~\ref{separate-kernel-execution-code}) implementation.

\subsubsection{Asymptotic expensive fused kernel} \label{taco_original_kernel}

The computation $A_{il} = \sum\nolimits {\sparse B_{ij}} \cdot C_{ik} \cdot D_{jk} \cdot E_{jl}$ can be fully realized using a nested loop iterator defined by all indices $i,j,k,\ \mbox{and}\ l$. 
The generalized way of producing kernels for a tensor multiplication of this kind in TACO is by generating an iteration graph (see Section~\ref{iteration_graph}). 
Since the iteration graph contains all the indices in a linear tree pattern, TACO generates a kernel as in Figure~\ref{fig:taco-fused-sddmm-spmm-kernel}, with time complexity of $O(nnz(B_{IJ})KL)$ due to the quadruple linearly nested loops (lines 1--6).

\subsubsection{Asymptotically inexpensive distributed kernels} \label{separate-kernel-execution-code}

However, the computation $A_{il} = \sum_{kj}\nolimits {\sparse B_{ij}} \cdot C_{ik} \cdot D_{jk} \cdot E_{jl}$ can be performed by evaluating two smaller kernels: sampled dense-dense matrix multiplication (SDDMM): 
$ {\sparse Y_{ij}} = \sum_{k}\nolimits {\sparse B_{ij}} \cdot C_{ik} \cdot D_{jk} $ followed by SpMM:
$ A_{il} = \sum_{j}\nolimits {\sparse Y_{ij}} \cdot E_{jl} $. 
As these separate kernels are triply nested loops (lines 2--6 in Figure~\ref{fig:sddmm-kernel} and 1--3 in Figure~\ref{fig:spmm-kernel}), they have lower asymptotic complexity.

Here, the Hadamard product, in SDDMM, results in ${\sparse Y_{ij}}$ matrix's sparse structure to be same as ${\sparse B_{ij}}$.
Therefore, the asymptotic complexity of performing two tensor computations with an intermediary matrix ${\sparse Y_{ij}}$ is $O(nnz(B_{IJ})(K+L))$. 
These separate kernels can be realized through loop distribution of the kernel from Section~\ref{taco_original_kernel}. 

Although we achieve a lower asymptotic complexity, we are using an intermediary tensor to 
pass values between SDDMM and SpMM, 
Hence, we miss the opportunity to exploit the temporal locality of the 
operation. 
The tensor contraction computed using linearly nested loops in Section~\ref{taco_original_kernel}. 
is expensive because of the high degree of nesting in the computation 
and the redundant duplicate computations, but 
may still be good for memory-constrained 
systems because the computation does not require any memory for storing intermediate results.

Using a temporary tensor to hold the result of the SDDMM operation is acceptable as long as the dimensionality of the index variables $i$ and $j$, and the density of the temporary tensor, are small. 
The code generation algorithm in TACO is limited to generating sequential code when the output tensor is of sparse format, (see $jY$ variable in Figure~\ref{fig:sddmm-kernel}). 
The kernel is sequential because the data format used to store the results of the computation limits random accesses. 
Here, the output of SDDMM operation is sparse (and the output from SpMM is dense) in which case we cannot parallelize the outermost loop of the SDDMM operation in separate kernel execution whereas the kernel in Figure~\ref{fig:taco-fused-sddmm-spmm-kernel} can be parallelized because the output of the combined kernel is dense. 
This is another valid reason to prefer the single kernel implementation despite its high asymptotic complexity.

\subsubsection{Fused kernel with low asymptotic complexity} \label{subsection:fused-kernel-code}

Since both the kernels $ {\sparse Y_{ij}} = \sum\nolimits {\sparse B_{ij}} \cdot C_{ik} \cdot D_{jk} $ 
in Figure~\ref{fig:sddmm-kernel} and 
$A_{il} = \sum\nolimits {\sparse Y_{ij}} \cdot E_{jl}$ in Figure~\ref{fig:spmm-kernel} have the same 
access patterns in their two outer-most loops, we can fuse them as shown in Figure~\ref{fig:SparFF-fused-kernel}, removing the use of the intermediary tensor to pass the values
between the two separate kernels as explained in Section~\ref{separate-kernel-execution-code}. 
This execution has a time complexity of $O(nnz(B_{IJ})(K+L))$, and at the same time removes the usage of 
a large tensor temporary by using an imperfectly nested loop structure (Lines 1--2,5 and 8 in Figure~\ref{fig:SparFF-fused-kernel}). 


Note that this partially-fused kernel provides the best of both worlds. Like the separate kernel approach, it has low asymptotic complexity. Like the fused kernel approach, it has good locality (since the temporaries only need to store data from the inner loops, their sizes much smaller and the reuse distances are reduced). 
Furthermore, because the outer loops of the partially fused are shared between both computations, and there is no longer a loop-carried dependence for SDDMM, the overall kernel can be parallelized in the same way as the kernel of Figure~\ref{fig:taco-fused-sddmm-spmm-kernel}.

\subsection{Our approach: \system}

While the schedule of computation in Figure~\ref{fig:SparFF-fused-kernel} provides both good asymptotic complexity and good locality, no existing system can automatically generate this type of schedule when generating code for sparse computations. TACO only handles ``linear'' iteration graphs that yield perfectly-nested loops, and hence cannot handle the partially-fused, imperfectly nested loop structure needed by our example. 
On the other hand, prior work on distribution and fusion for tensor computations~\cite{saday1}, can support this type of code structure only for operations on dense tensors.

\system provides mechanisms for generating the code in Figure~\ref{fig:SparFF-fused-kernel} from a high level representation of the computation as well as scheduling directives that inform the structure of the code. 
We introduce several components to perform this code generation and Section~\ref{method} discuss them in detail. 

\begin{enumerate}[align=left,leftmargin=*]
\item We introduce a new representation called a {\em branched iteration graph} that allows the representation of partially-fused iteration structures, where some loops in a loop nest are common between computations and others are separate. Hence, this graph represents imperfect nesting. We carefully place constraints on these graphs to ensure that the requirements of nested iteration over sparse structures are met. The branched iteration graph is described in more detail in Section~\ref{algorithm}.
\item We introduce new scheduling primitives for loop distribution and fusion that allow programmers to {\em generate} the branched iteration graph by applying scheduling transformations to linear TACO iteration graph. We describe the primitives and describe how they systematically transform a branched iteration graph in Section~\ref{support_existing_transformations}.
\item We adapt TACO's code generation strategies to the branched iteration graph, allowing \system to generate sparse iteration code for tensor kernels that have had our distribution and fusion transformations applied to them. We discuss code generation in Section~\ref{code_generation}.
\end{enumerate}

\section{Detailed Design} \label{method}

This section describes the key components of \system. Section~\ref{subsec:representation} describes \system's new branched iteration graph representation. Section~\ref{subsec:transformation} shows how partial fusion is represented through iteration graph transformations. Section~\ref{subsec:directives} explains how scheduling directives can guide partial fusion while still composing with TACO's existing scheduling language. Finally, Section~\ref{subsec:codegen} explains how \system generates code.

\begin{figure*}[ht]
    \vspace{-1em}
    \centering
        \hfill
        \hfill
        \begin{subfigure}[t]{0.14\textwidth}
            \centering
            \includegraphics[width=0.85\linewidth]{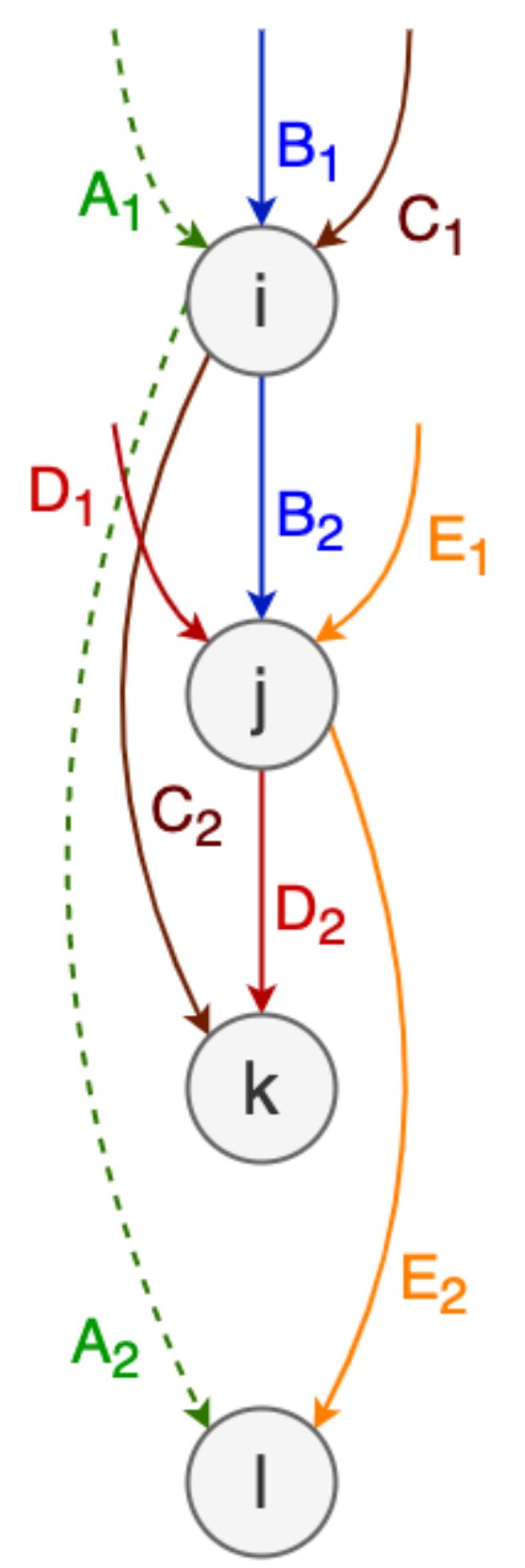}
            \caption{Original kernel}
            \vspace{-5pt}
            \label{fig:sddmm-spmm-original}
            \Description[Original iteration graph of sddmm spmm operation]{<long description>}
        \end{subfigure}%
        \hfill
        \begin{subfigure}[t]{0.13\textwidth}
            \centering
            \includegraphics[width=0.85\linewidth]{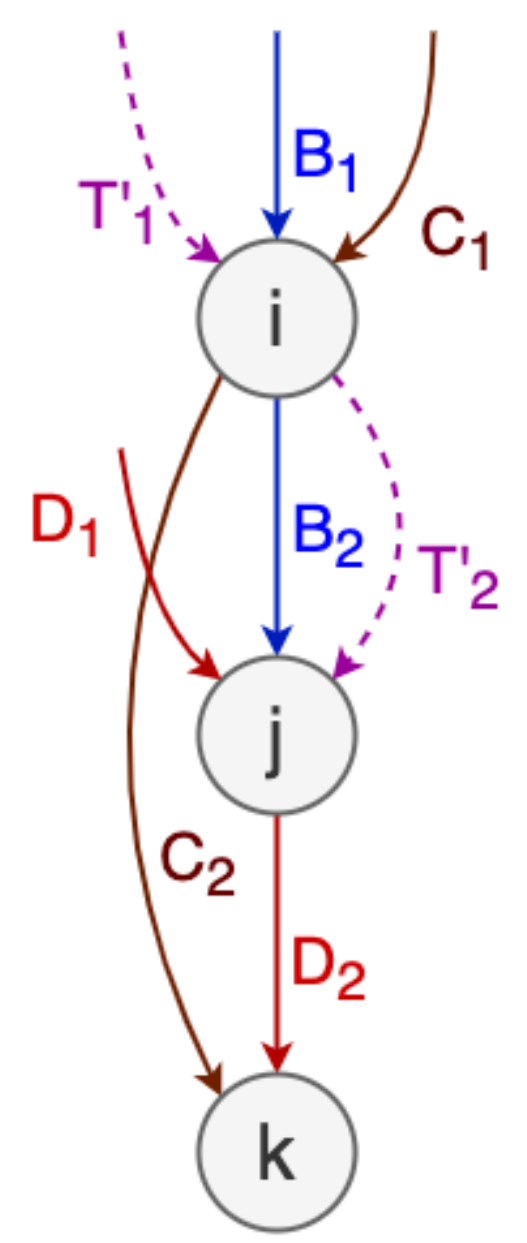}
            \caption{SDDMM}
            \vspace{-5pt}
            \label{fig:sddmm-spmm-removed}
            \Description[SDDMM iteration graph recovered from the iteration graph of the combined iteration graph of SDDMM and SpMM operation]{<long description>}
        \end{subfigure}%
        \hfill
        \begin{subfigure}[t]{0.13\textwidth}
            \centering
            \includegraphics[width=0.85\linewidth]{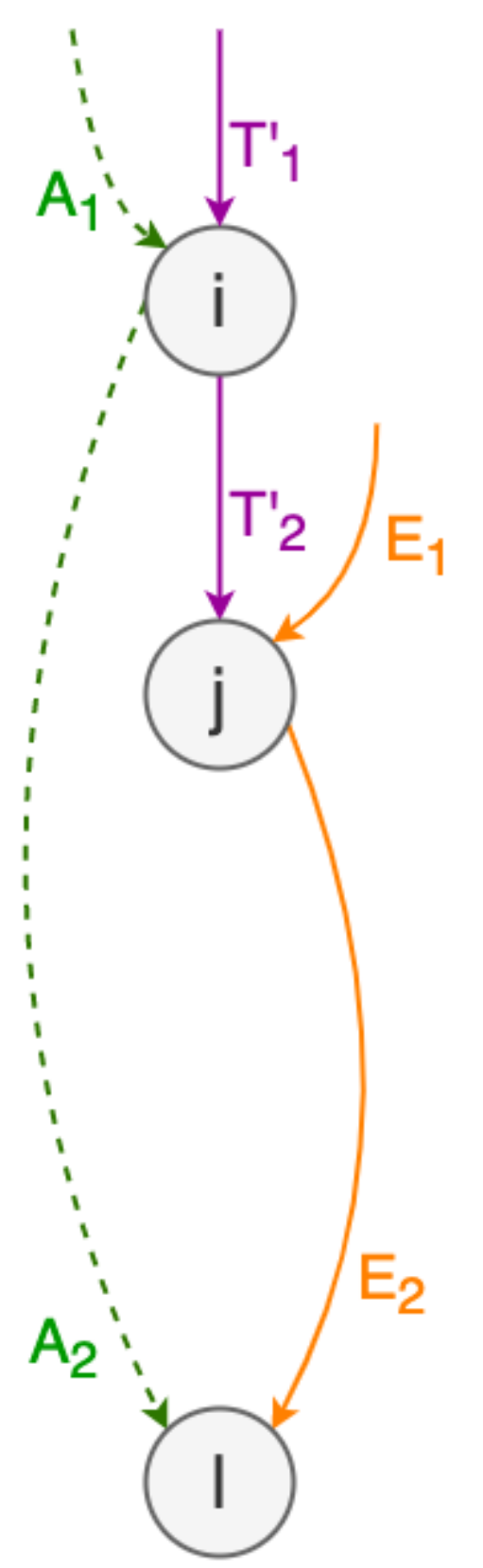}
            \caption{SpMM}
            \vspace{-5pt}
            \label{fig:sddmm-spmm-add-back}
            \Description[Iteration graph of the SpMM operation recovered from the combined kernel]{<long description>}
        \end{subfigure}%
        \hfill
        \begin{subfigure}[t]{0.26\textwidth}
            \centering
            \includegraphics[width=0.85\linewidth]{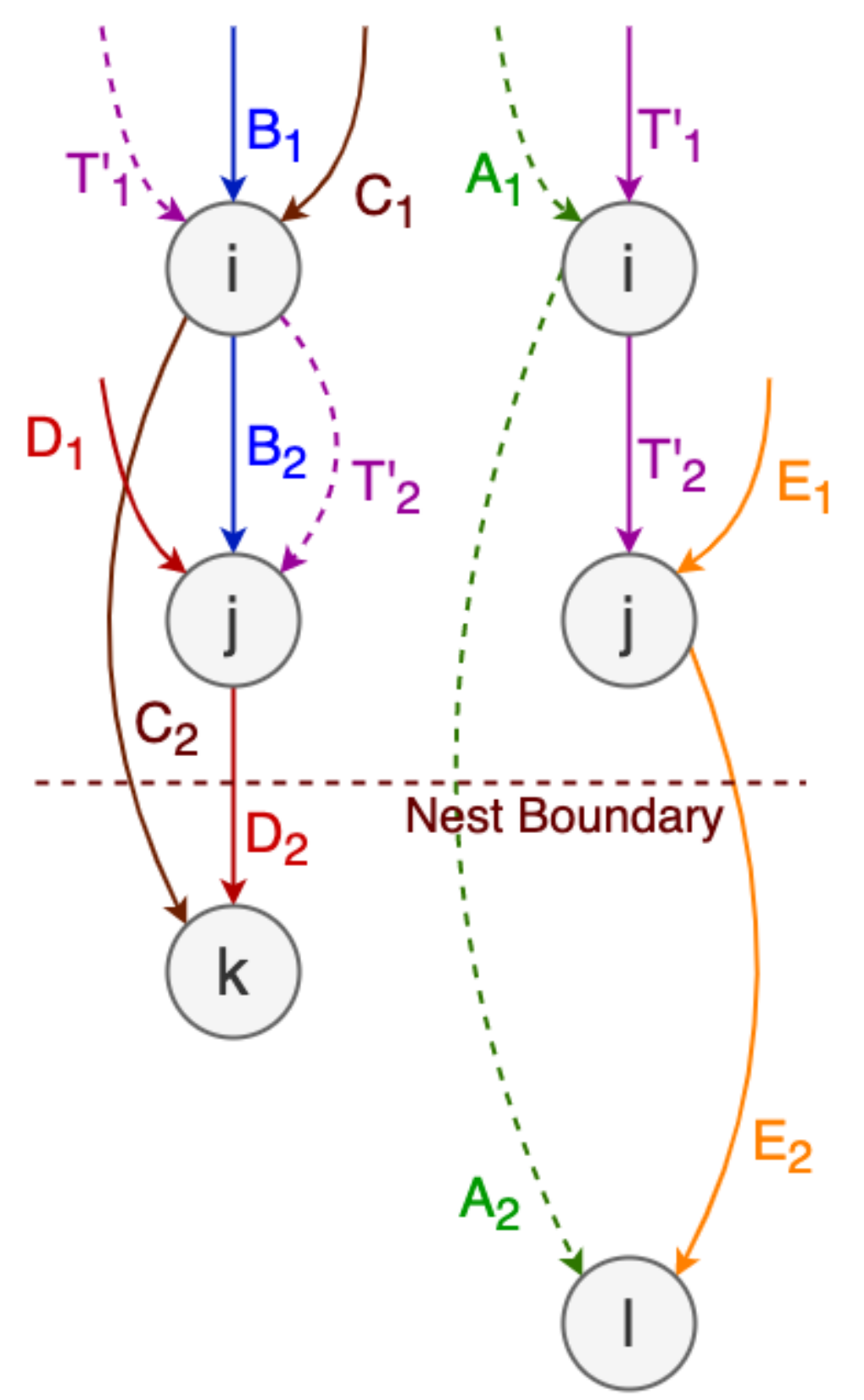}
            \caption{Producer/Consumer kernels}
            \vspace{-5pt}
            \label{fig:sddmm-spmm-combined}
            \Description[Reasoning about the producer and consumer iteration graphs]{<long description>}
        \end{subfigure}%
        \hfill
        \begin{subfigure}[t]{0.26\textwidth}
            \centering
            \includegraphics[width=0.85\linewidth]{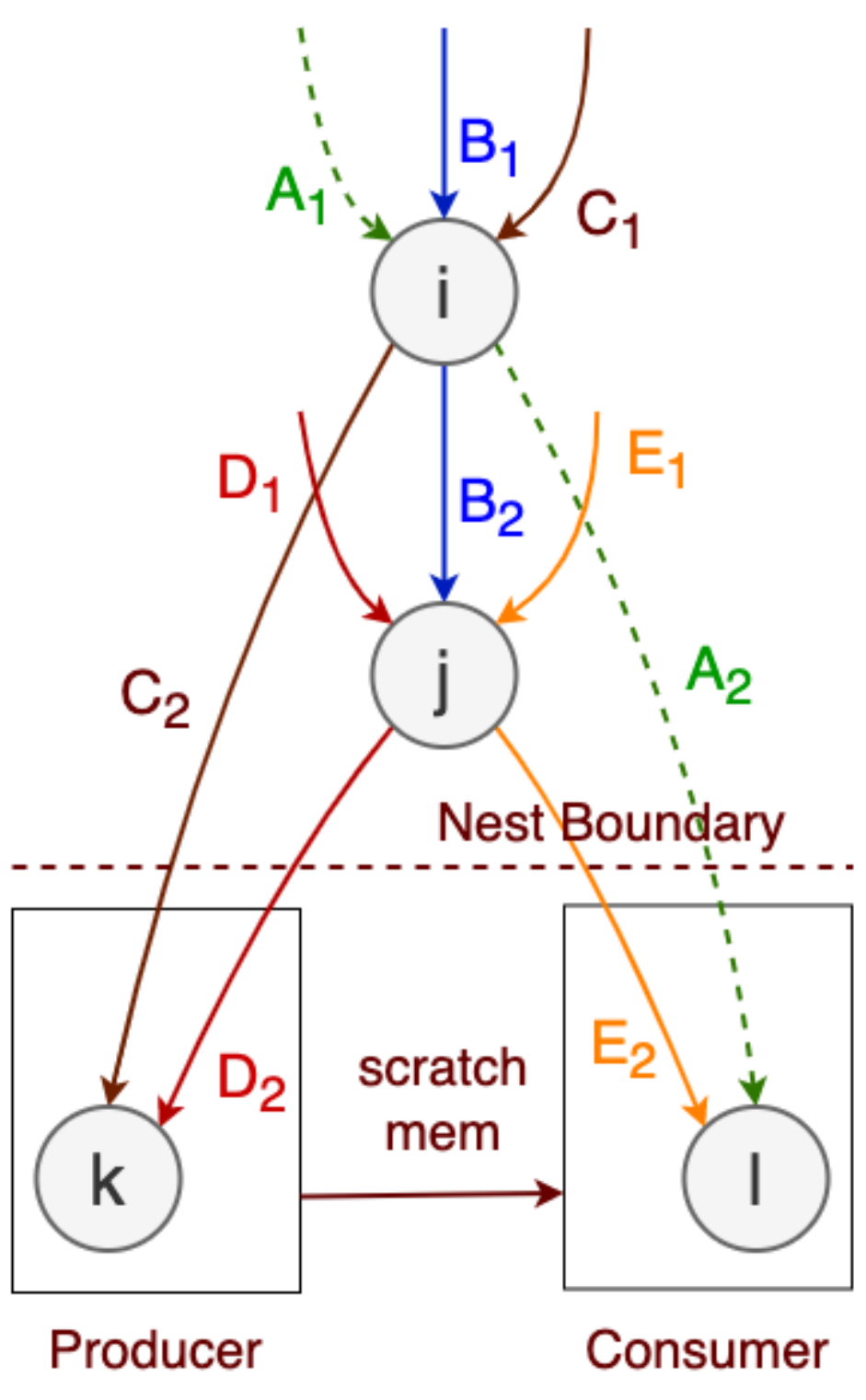}
            \caption{Fused kernel}
            \vspace{-5pt}
            \label{fig:sddmm-spmm-branched}
            \Description[Branched iteration graph for the SDDMM, SpMM operation]{<long description>}
        \end{subfigure}%
        \hfill
        \hfill
    \caption{loopfuse transformation performed on $ A_{il} = \sum\nolimits_{jk} {\sparse B_{ij}} C_{ik} D_{jk} E_{jl} $}
    \label{fig:sddmm-spmm-all-graphs}
    \vspace{-1.0em}
\end{figure*}

\subsection{Representation}
\label{subsec:representation}

\system uses a {\em branched iteration 
graph} to represent sparse tensor algebra kernels, 
which is an extension to the concrete index notation described in~\cite{kjolstad:2018:workspaces}. A branched iteration graph can be understood as 
an iteration graph with branches in index access patterns. By transforming the linear 
index tree iteration graph generated by TACO to a branched iteration graph in the 
context of tensor multiplication, we try to 
remove the asymptotic complexity that arises from perfectly linearly nested loops in dense/sparse 
iterations.

\begin{definition}
    A branched iteration graph is a directed graph $G = (V, G_{p}, G_{c}, P)$, where $V$ is a set of {\em unbranched} indices, organized as a sequence starting from the root of the iteration graph that then has two children graphs,
    $G_{p}$ (producer) and $G_{c}$ (consumer), that define the two branches of $G$, where $G_p$ and $G_c$ themselves are branched iteration graphs, such that there is a {\em dependence edge} from $G_{p}$ to $G_{c}$ and a \textit{boundary} between $V$ and ($G_c, G_p$). 
    The {\em dependence edge} tracks the common set of indices in $G_{p}$ and $G_{c}$. 
    $P = {p_1, p_2, \ldots, p_n}$ defines the set of tensor paths, a tuple of indices associated with a particular tensor variable.
\end{definition}

Intuitively, where a TACO iteration graph corresponds to a perfectly-nested loop where the order of the vertices in the graph corresponds to the nesting order of the loops, a branched iteration graph represents an imperfectly nested loop. $V$ corresponds to the common outer loops, just as in a TACO graph, while $G_p$ and $G_c$ correspond to the inner loop nests (which can themselves be imperfectly nested).
For example, in Figure~\ref{fig:sddmm-spmm-branched}, $V$ refers to the set of indices \{$i,j$\}, and $G_c$, $G_p$ refer to the boxes \textit{Producer} and \textit{Consumer}, respectively.


\subsection{Branched Iteration Graph Transformation} \label{algorithm}\label{subsec:transformation}

In Section~\ref{motivating_example} we saw how we could perform loop fusion or distribution for a sparse tensor algebra computation.
We recognize this pattern in index traversal and exploit it to generate the branched iteration graph.
We name this pattern recognition algorithm \emph{fusion after distribution} because it proceeds in two steps as described in Algorithm~\ref{algorithm:sparsefd}:
(i) distributing the perfectly-nested indices in the iteration graph, and then 
(ii) fusing the common indices.
\begin{algorithm}
	\caption{Loop fusion after distribution}
    \label{algorithm:sparsefd}
    
    \begin{algorithmic}[1]
    \Require{Topologically Ordered Iteration Graph $G_{I}=(I_{G},P)$}
    \Require{Index Expression $Expr$: $A_{out} = A_{1}*A_{2}*...*A_{n}$}
    \Require{Bool $recursive$}
    \Ensure{Branched Iteration Graph $G'_{I}$}
    \vspace{0.5em}
    \State fusible = isFusible($G_{I}$) \label{lst:line:is_fusible}
    \IIf{!fusible} \textbf{return} $G_{I}$
    \State $P_{T'} = P_{A_{out}} - P_{A_{n}}$  \Comment{index path for temporary tensor $T'$}
    \State $G_{I-Producer}, ProducerExpr_{temp} := T'(P_{T'}) = Expr \setminus A_{n}$
    \State $G_{I-Consumer}, ProducerExpr_{temp} := A_{out} = T'(P_{T'}) * A_{n}$
    \If{recursive} \label{lst:line:recursive_start}
        \State $G_{I-Producer} = recursiveCall(G_{I-Producer},$ \WRP $ProducerExpr_{temp}, recursive)$

    \EndIf \label{lst:line:recursive_end}
    \State $List_{I-Producer} = GetIndices(G_{I-Producer})$ \label{lst:line:common_prefix_start}
    \State $List_{I-Consumer} = GetIndices(G_{I-Consumer})$
    \State \textbf{Define}: $I_{sharable} = \emptyset$ \label{lst:line:sharable_start}
    \For{Each $i \in I_{G}$} 
        \If{$i \in List_{I-Producer} \textbf{ and } i \in List_{I-Consumer}$}
            \State $I_{sharable} = I_{sharable} \cup i$
        \Else \quad break;
        \EndIf
    \EndFor \label{lst:line:sharable_end}
    \State \textbf{Define}: $I_{fusable} = \emptyset$ \label{lst:line:fusible_start}
    \For{$i \gets 1$ to $N$} 
        \If{$i\not\in I_{sharable} \textbf{ and } i \in List_{I-Producer} \textbf{ and }$ \WRP $i \in List_{I-Consumer}$}
            \State $I_{fusable} = I_{fusable} \cup i$
        \Else \quad break;
        \EndIf
    \EndFor \label{lst:line:fusible_end}
    \State \textbf{Define}: $T(P_{I_{fusable}})$ \label{lst:line:temp_define}
    \State $ProducerExpr := T(P_{I_{fusable}}) = T'(P_{T'})$ \label{lst:line:producer_with_temporary}
    \State $ConsumerExpr := A_{out} = T(P_{I_{fusable}}) * A_{N}$ \label{lst:line:consumer_with_temporary}
    \State \Return{$GraphRewrite(G_{I}, I_{sharable},$\WRP$ProducerExpr, ConsumerExpr)$}
    \end{algorithmic}
\end{algorithm}

\paragraph{Topologically sorted iteration graph.}
The iteration graph in Figure~\ref{fig:sddmm-spmm-original} relates to the index expression 
$ A(i,l) = {\sparse B(i,j)} * C(i,k) * D(j,k) * E(j,l) $, where B is sparse. 
We denote this kernel as <SDDMM, SpMM>. The indices here are 
topologically ordered such that the ordering of the indices are constrained by the 
sparsity patterns of the sparse tensors. The ordering 
$ i \rightarrow j \rightarrow k \rightarrow l $ would be consistent 
with the access patterns of all the tensors 
$i \rightarrow l$ in A, $i \rightarrow j$ in B, $i \rightarrow k$ in C, $j \rightarrow k$ in D,
and $j \rightarrow l$ in E. However, there should be a hard ordering imposed on $i$ and $j$ index 
variables because $j$ cannot be accessed without accessing $i$ first. 
The index access patterns of the tensor access variables are marked in the graph as paths. $A_{1}$ 
denotes the first access dimension of the $\dense{A}$ and $A_{2}$ denotes the second access dimension 
of the $\dense{A}$ tensor. The fusion algorithm requires the iteration graph in 
Figure~\ref{fig:sddmm-spmm-original} and the tensor index notation expression $ A(i,l) = {\sparse B(i,j)} * C(i,k) * D(j,k) * E(j,l) $.
We identify the iteration graph as fusible if there are indices that are only present in the last tensor and the output tensor in the tensor expression (line~\ref{lst:line:is_fusible} of the Algorithm~\ref{algorithm:sparsefd}).

\paragraph{Distribution into two kernels.}
The description of the tensor kernel above captures all the information of performing 
kernel executions 
SDDMM: ${\sparse T'(i,j)} = {\sparse B(i,j)} * C(i,k) * D(j,k) $ and 
SpMM: $A(i,l) = {\sparse T'(i,j)} * E(j,l) $ sequentially. (Notice that 
separation of kernels requires a temporary matrix $T'$)
Therefore, we can recover the separate 2 smaller kernels that would yield the same 
result given the larger tensor expression. 
We denote the first kernel as the producer and the second kernel as the consumer.
 To find these separate smaller kernels, we need to remove the last 
tensor $E(j,l)$ from the original expression.
Line 8 of the Algorithm~\ref{algorithm:sparsefd} creates the producer index expression and iteration graph 
for the tensor computation performed first (SDDMM in our running example) by removing the 
last tensor from the original expression, 
and then line 9 of the Algorithm~\ref{algorithm:sparsefd} creates the consumer index expression and 
iteration graph for the tensor computation that is performed second (SpMM in our 
running example) by adding it back to the producer's expression. 
These 2 separate kernels would have iteration graphs shown in Figures~\ref{fig:sddmm-spmm-removed} 
and~\ref{fig:sddmm-spmm-add-back} respectively. 
We perform this recovery of the two separate operations in order to identify the fusible 
and shared indices between two separate tensor operations as we will further explain in a 
next paragraph. 

\paragraph{Fusing common loops.} 
Once we have the iteration graphs for the separate kernels we reason about them together (See~\ref{fig:sddmm-spmm-combined}).
We reason that 
both the sparse iterations need to iterate through the space using index variables $i$ and $j$. 
Also, iteration space defined by the index $k$ is iterated only by the SDDMM operation, and the iteration 
space defined by the index $l$ is only iterated by the SpMM operation. 
But those iterations over index $k$ and $l$ need to happen one after the other. 
The producer-consumer dependence must be satisfied 
such that the values consumed by the consumer must have been produced by the producer before its use. 
The values shared between the producer and consumer can be stored in an intermediate scratch memory. 
Furthermore, the comparison of the two graphs, the producer 
graph and the consumer graph, helps identify the indices that can and cannot be shared among the iterations.

The producer graph in Figure~\ref{fig:sddmm-spmm-removed} and the consumer 
graph in Figure~\ref{fig:sddmm-spmm-removed} 
have a common prefix defined by some indices in their iteration graphs. We run a prefix match to identify the shared 
indices by the two kernels (lines~\ref{lst:line:common_prefix_start}--\ref{lst:line:sharable_end} of the Algorithm~\ref{algorithm:sparsefd}), 
in Figure~\ref{fig:sddmm-spmm-combined}. We see that both $i$ and $j$ indices are shared, and 
the other variables are not shared. 
The addition of indices $i$ and $j$ to the set of sharable indices 
is described in lines~\ref{lst:line:sharable_start}--\ref{lst:line:sharable_end} of the Algorithm~\ref{algorithm:sparsefd}. 
We identify this point as a \emph{nest boundary} in the iteration graph~\ref{fig:sddmm-spmm-combined}, and denote the indices above the \emph{nest boundary} as fusible.
The final output of executing the {\em fusion after distribution algorithm} is a branched iteration graph.
Therefore, if the algorithm is applied recursively (see the kernel <SDDMM, SpMM, GEMM> in the benchmark Section~\ref{benchmarks}) on the producer (lines~\ref{lst:line:recursive_start}--\ref{lst:line:recursive_end} of the Algorithm~\ref{algorithm:sparsefd}), our algorithm can still match the prefix even if the producer graph is already branched.

\paragraph{Materializing temporary variables.}
The next step of the Algorithm is to identify the indices that cannot be fused as outermost 
loops but are common to the 
producer and the consumer. In Figure~\ref{fig:sddmm-spmm-combined} we see that there are no common variables below the \emph{nest boundary}.
The variables that are below the \emph{nest boundary} line and common to both the producer and consumer define the dimensions of the temporary variable that is shared between them.
For the case of <SDDMM, SpMM> described in Figure~\ref{fig:sddmm-spmm-combined}, since no indices are 
common below the \emph{nest boundary} line, we can define the temporary as a scalar.
However, for the same case of <SDDMM, SpMM> described in 
Section~\ref{support_existing_transformations}, where transpose of $D$ is used to define the computation, we can see that index $j$ is a common index below the \emph{nest boundary} line.
Therefore, the algorithm 
defines a temporary vector bounded by the size of the index $j$, 
Lines~\ref{lst:line:fusible_start}-\ref{lst:line:fusible_end} of the Algorithm~\ref{algorithm:sparsefd} explain how we perform the identification of the common indices below the \emph{nest boundary}, and line~\ref{lst:line:temp_define} defines this temporary variable.

\paragraph{Rewrite the iteration graph.}
After we find the fusible indices, shared indices and define the temporary variable, we define the producer expression and consumer expression using the temporary variable that is shared between the producer and the consumer (lines~\ref{lst:line:producer_with_temporary},~\ref{lst:line:consumer_with_temporary} of the Algorithm~\ref{algorithm:sparsefd}).
Then, we rewrite the iteration graph to 
model this behavior with the temporary variable, the producer and the consumer (see Figure~\ref{fig:sddmm-spmm-combined}) which would eventually generate the code shown in Figure~\ref{fig:SparFF-fused-kernel} for our running example. 

\subsection{Scheduling} \label{scheduling}

In this section we describe, (1) the invocation of scheduling transformation and (2) the impact it has on the space of possible schedules.

\subsubsection{Scheduling Directive} \label{support_existing_transformations}\label{subsec:directives}

\begin{figure}[!t]
    \centering
    \begin{minipage}[b]{.33\columnwidth}
    \begin{subfigure}{\linewidth}
    \vspace{-5em}
    \includegraphics[scale=0.40]{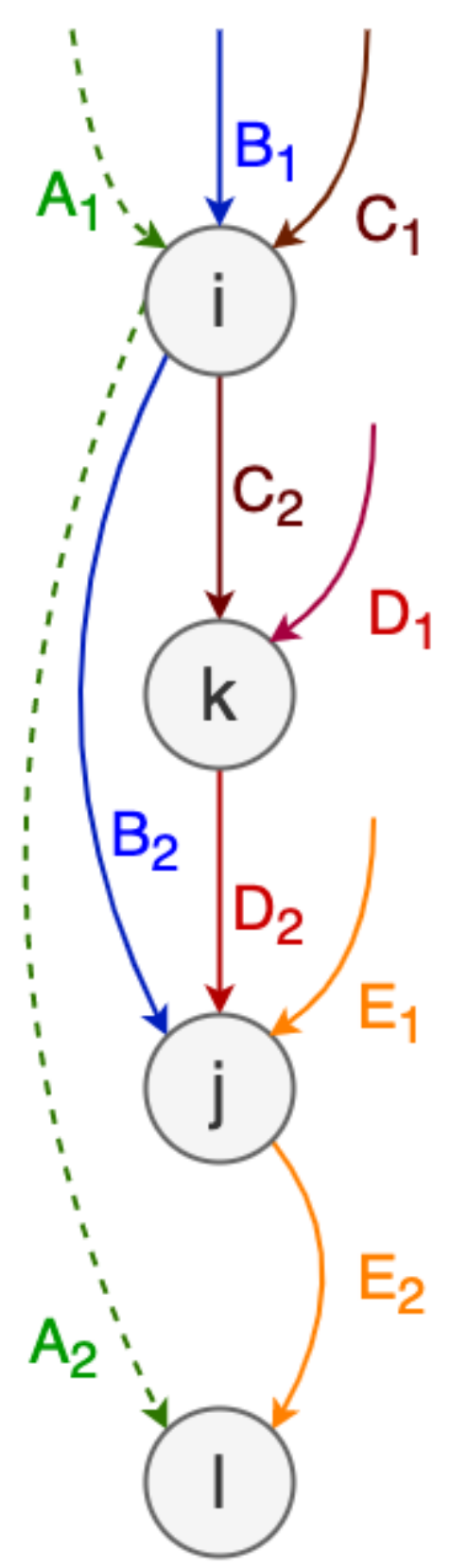}
    \caption{Original kernel}
    \label{fig:sddmm-spmm-reorder-original}
    \end{subfigure}
    \end{minipage}
    \begin{minipage}[b]{.50\columnwidth}
    \begin{subfigure}{\linewidth}
    	\centering
    	\includegraphics[width=.9\linewidth]{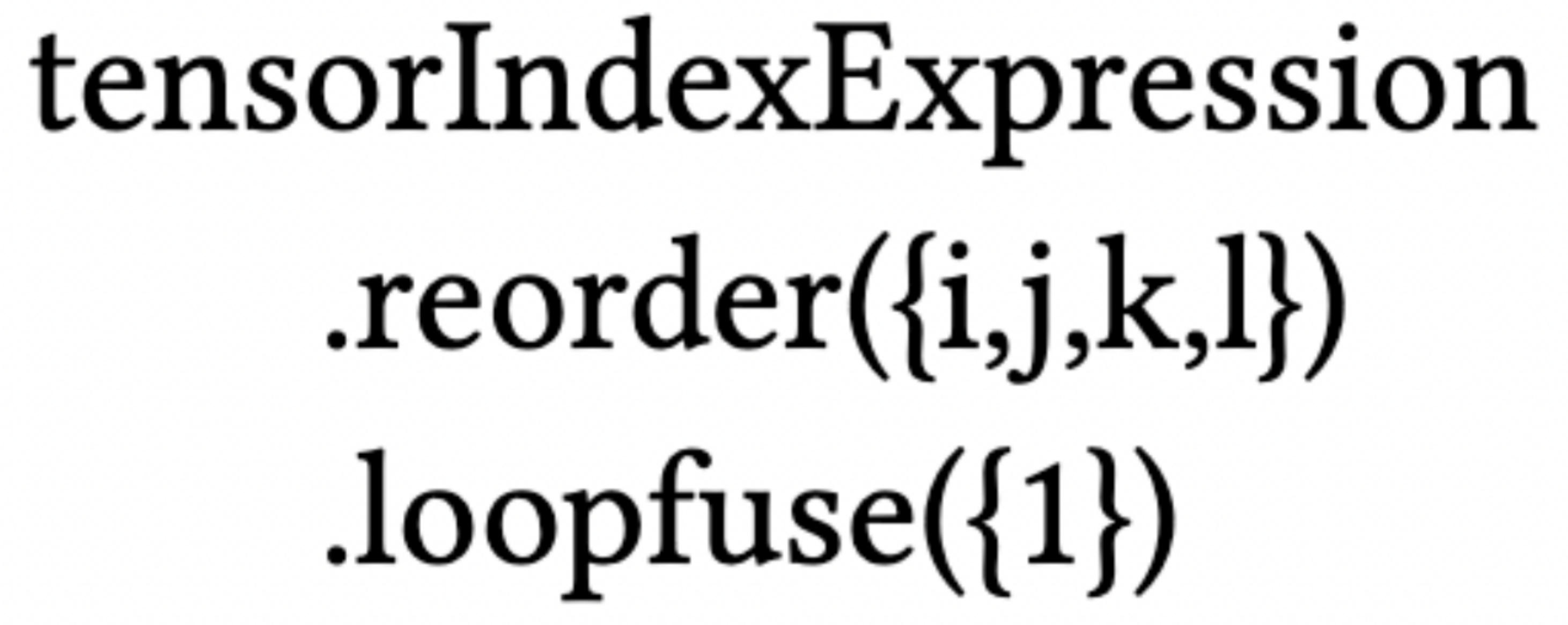}
    	\caption{Scheduling directives}
    	\label{fig:scheduling-primitive}
    \end{subfigure}
    \begin{subfigure}{\linewidth}
    	\centering
    	\includegraphics[width=.9\linewidth]{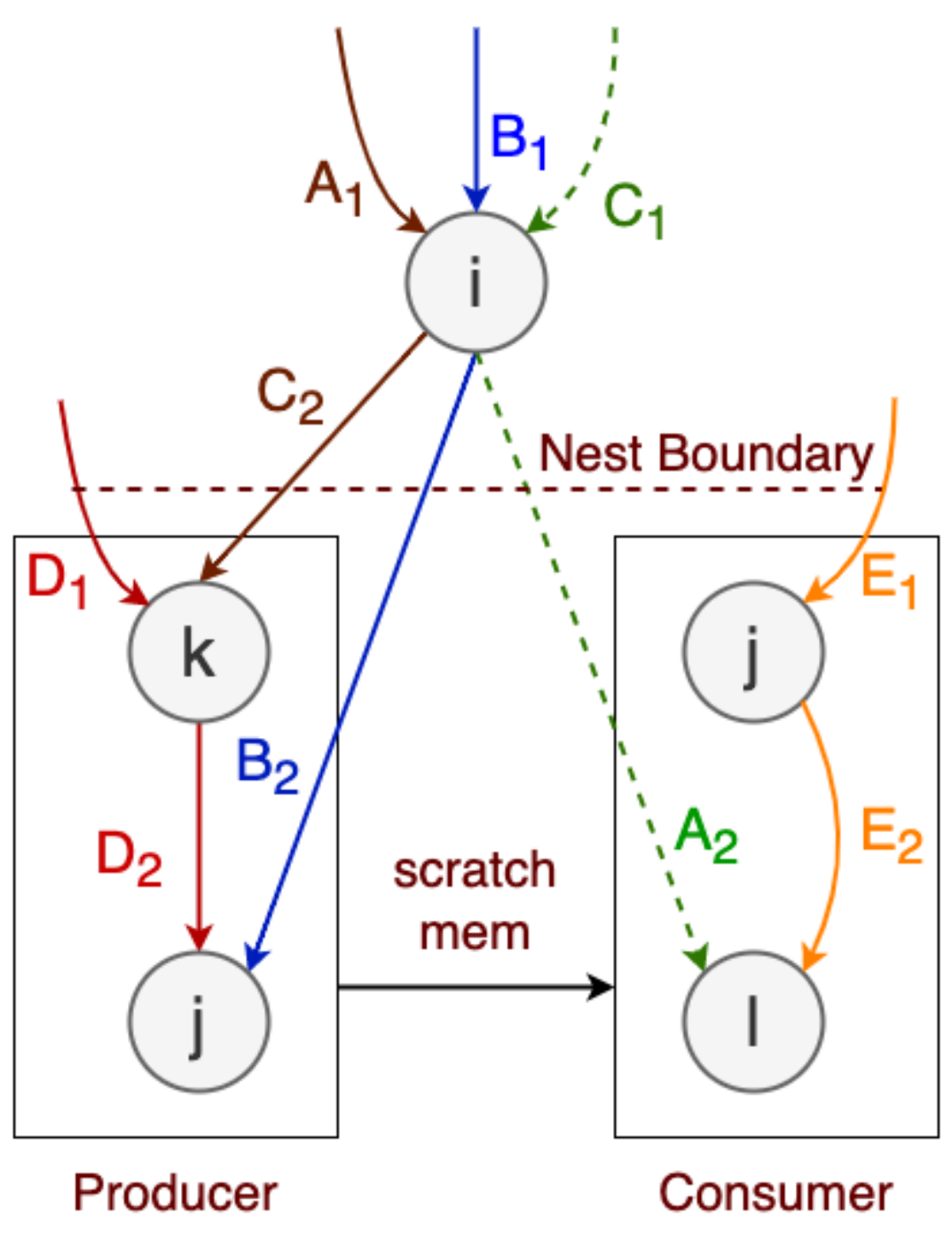}
    	\caption{Transformed kernel}
    	\label{fig:sddmm-spmm-reorder-branched}
    \end{subfigure}
    \end{minipage}
    \vspace{-1.0em}
    \caption{The \code{loopfuse} transformation performed on $A_{il} = \sum\nolimits_{jk} B_{ij} C_{ik} D_{kj} E_{jl}$.}%
    \label{fig:sddmm-spmm-reorder}
    \vspace{-1.5em}
\end{figure}

\system introduces a new scheduling
directive to TACO. 
The user can call the \code{loopfuse} scheduling transformation as shown in Figure~\ref{fig:scheduling-primitive} with other scheduling directives. 
Here, \textbf{1} refers to applying the algorithm once. 
By passing \textbf{2} or a higher number, the algorithm can be applied recursively.

Sometimes it is necessary to combine \code{loopfuse} with other TACO scheduling directives. Hence, it is important that our new directive compose with the existing scheduling language.
For example, 
applying Algorithm~\ref{algorithm:sparsefd} to the tensor expression 
$A(i,l) = {\sparse B(i,j)}*C(i,k)*D(k,j)*E(j,l)$ would not yield the code in Figure~\ref{fig:SparFF-fused-kernel} 
by default because now the access pattern of the D matrix is different since we are using 
the transpose of $D$ for this example. This difference results 
in a different iteration graph as shown in Figure~\ref{fig:sddmm-spmm-reorder-original}, 
because now the iteration 
graph needs to preserve the ordering of $i \rightarrow j$ for B, $i \rightarrow k$ for D, 
$k \rightarrow j$ for D, $j \rightarrow l$ for E, and $i \rightarrow l$ for A, with a hard ordering 
of $i \rightarrow j$ because B is a sparse matrix.
Applying the \emph{fusion after distribution} algorithm would result in an iteration graph as depicted in 
Figure~\ref{fig:sddmm-spmm-reorder-branched}. 

However, since D is dense, there is no 
hard constraint on the ordering of indices $k$ and $j$. Therefore, to arrive at the code in Figure~\ref{fig:SparFF-fused-kernel}, a loop reordering can be performed before the loopfuse 
scheduling directive (See Figure~\ref{fig:scheduling-primitive}). 

The \emph{nest boundary} between the branching point in the iteration graph constrains loop reordering between the \emph{nest boundary}.
But loop reordering can be still allowed in indices within \emph{nest boundaries}.

\subsubsection{Scheduling Space}

In our current implementation, if two kernels have $n$ common outer loops, we fuse them all.
However, if loop reordering is possible, and we choose only certain loops to be fused, then there are $2^n$ possible schedules to start with, given that there are $n$ number of fusible loops (i.e. $n$ number of common iterators in the two kernels). 
This is an upper bound without considering any constraints of sparse access patterns. 
Reordering of inner loops can be performed after fusion, and other scheduling directives (split, parallelization, etc.) can be applied separately, giving more scheduling opportunities with imperfect nesting. This scheduling space is obviously very large, so smart strategies for searching that space is a promising avenue for future work.

\subsection{Code Generation} \label{code_generation}\label{subsec:codegen}

We carefully redesigned intermediate representation (IR) in TACO to support the branched iteration graph and manage temporaries such that code generation backend does not require any changes. 
We rewrite the graph loop structure with \code{where} statements 
defining a producer-consumer relationship. 
This placement of temporaries for the producer-consumer relationship and the change of iteration graph explained in Section~\ref{algorithm}  
preserves all the attributes that are necessary for TACO 
code generation backend.

In TACO each index in the iteration graph is converted to one or more loops to iterate through dense loops or co-iterate over the levels of sparse data formats. 
An iteration lattice~\cite{kjolstad:2017:taco} is used to co-iterate through the intersections of the sparse dimensions which results in a single for-loop, single while-loop or multiple while-loops.

\section{Implementation} \label{implementation}

We implement the branch iteration graph transformation described in Section~\ref{method} on top of the TACO ~\cite{kjolstad:2017:taco} intermediate representation (IR). 
Furthermore, we introduce a new scheduling directive to separate it from the algorithmic language and to provide the scheduling language with more opportunities to generate more (performant) schedules. 

We change the iteration graph~\cite{kjolstad:2017:taco} and use 
the concrete index notation~\cite{kjolstad:2018:workspaces} to introduce intermediate temporaries that are shared between the producer and the consumer. 
We implement a \emph{nest boundary} between the fused loops and shared index loops to constrain performing loop reordering transformations between them.
In our running example, the user cannot interchange loops with an outer level, once the distribution operation is performed.

This new transformation can be used in the context of tensor multiplication. 
Hence, it does not generalize to tensor expressions with tensor additions. 
We limit the number of tensors and index variables removed from the index expression, to identify the producer and consumer graphs, per iteration to one. 
We believe that the algorithm could be generalized to support fusion of indices shared between multiple tensors which would be able to support high order tensors and complex tensor contractions. 
\section{Evaluation} \label{evaluation}

We compare \system to two other techniques:

\noindent
\textbf{TACO Original.}
Given a large combined index expression containing multiple smaller index expressions, the code generated by TACO has a perfectly nested loop structure with at least one loop per each index variable in the index expression. 
We refer to this version as {\em TACO Original}. 

\noindent
\textbf{TACO Separate.}
In some cases, the asymptotic complexity of {\em TACO Original} can be reduced 
by manually separating a larger index expression 
into multiple smaller index expressions by using temporary 
tensors to store the intermediate results. 
We refer to this version as {\em TACO Separate}.
When there are multiple ways to break down 
the computation into smaller kernels, we evaluate all those combinations and report the best execution time.

\begin{figure*}[ht]
    \centering
    \begin{subfigure}[t]{0.43\textwidth}
        \centering
        \includegraphics[width=0.96\linewidth]{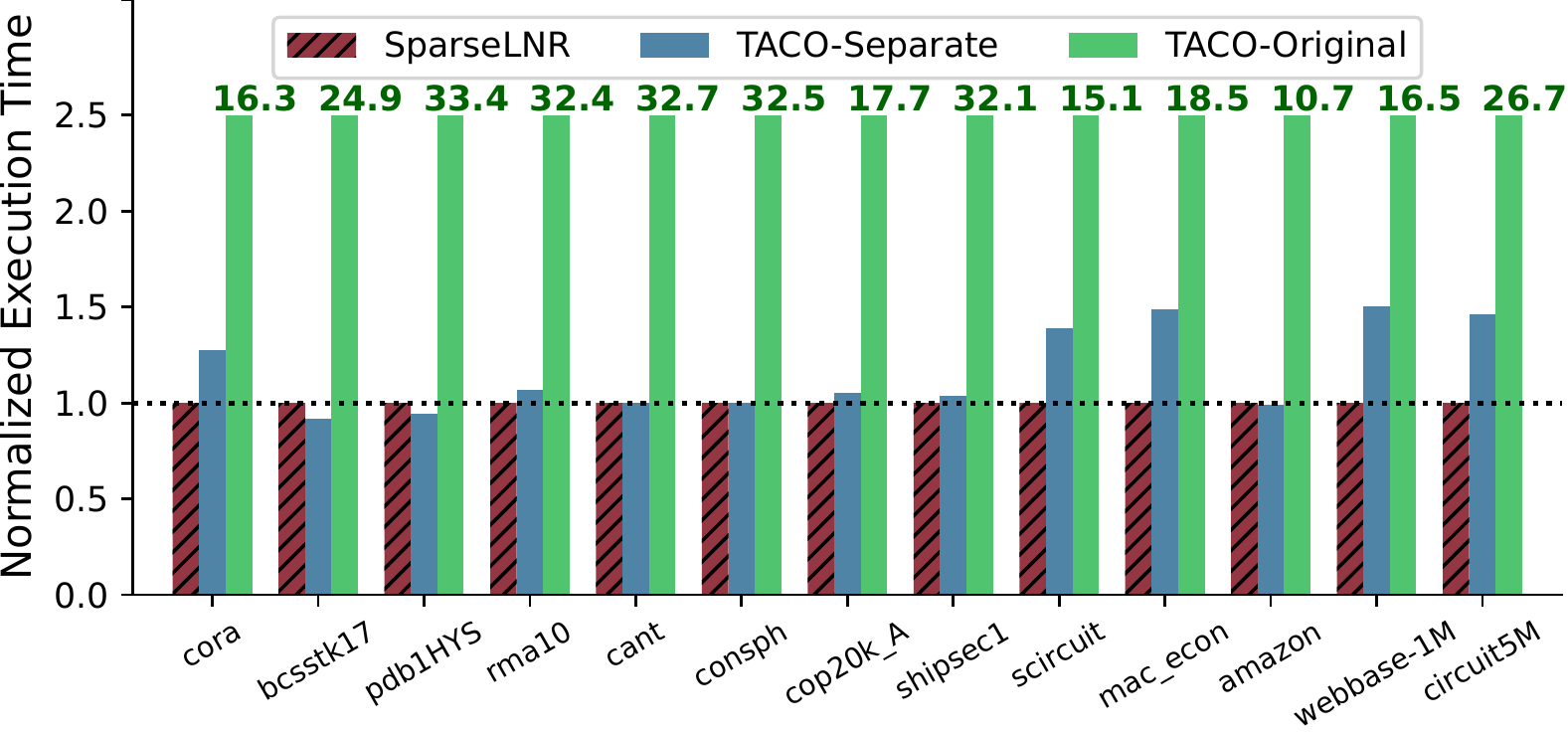}
        \vspace{-4pt}
        \caption{Single-threaded <SDDMM, SpMM>}
        \vspace{-10pt}
        \label{fig:sddmm-spmm-single-thread}
        \Description[sddmm-spmm original iteration graph]{<long description>}
    \end{subfigure}%
    \hspace{2em}%
    \begin{subfigure}[t]{0.43\textwidth}
        \centering
        \includegraphics[width=0.96\linewidth]{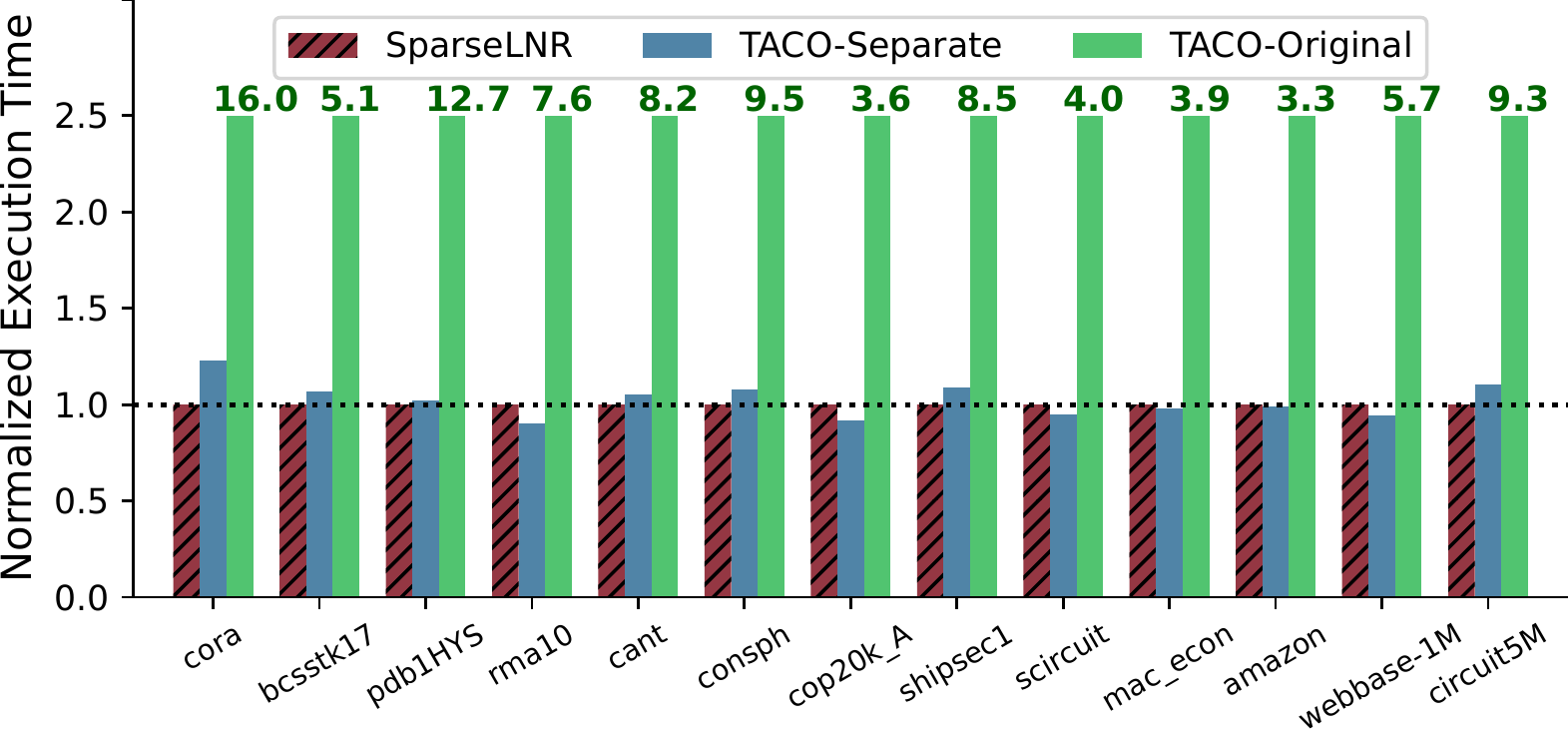}
        \vspace{-4pt}
        \caption{Multi-threaded <SDDMM, SPMM>}
        \vspace{-10pt}
        \label{fig:sddmm-spmm-multi-thread}
        \Description[sddmm-spmm original iteration graph]{<long description>}
    \end{subfigure}%
    \vskip\baselineskip
    \begin{subfigure}[t]{0.43\textwidth}
        \centering
        \includegraphics[width=0.96\linewidth]{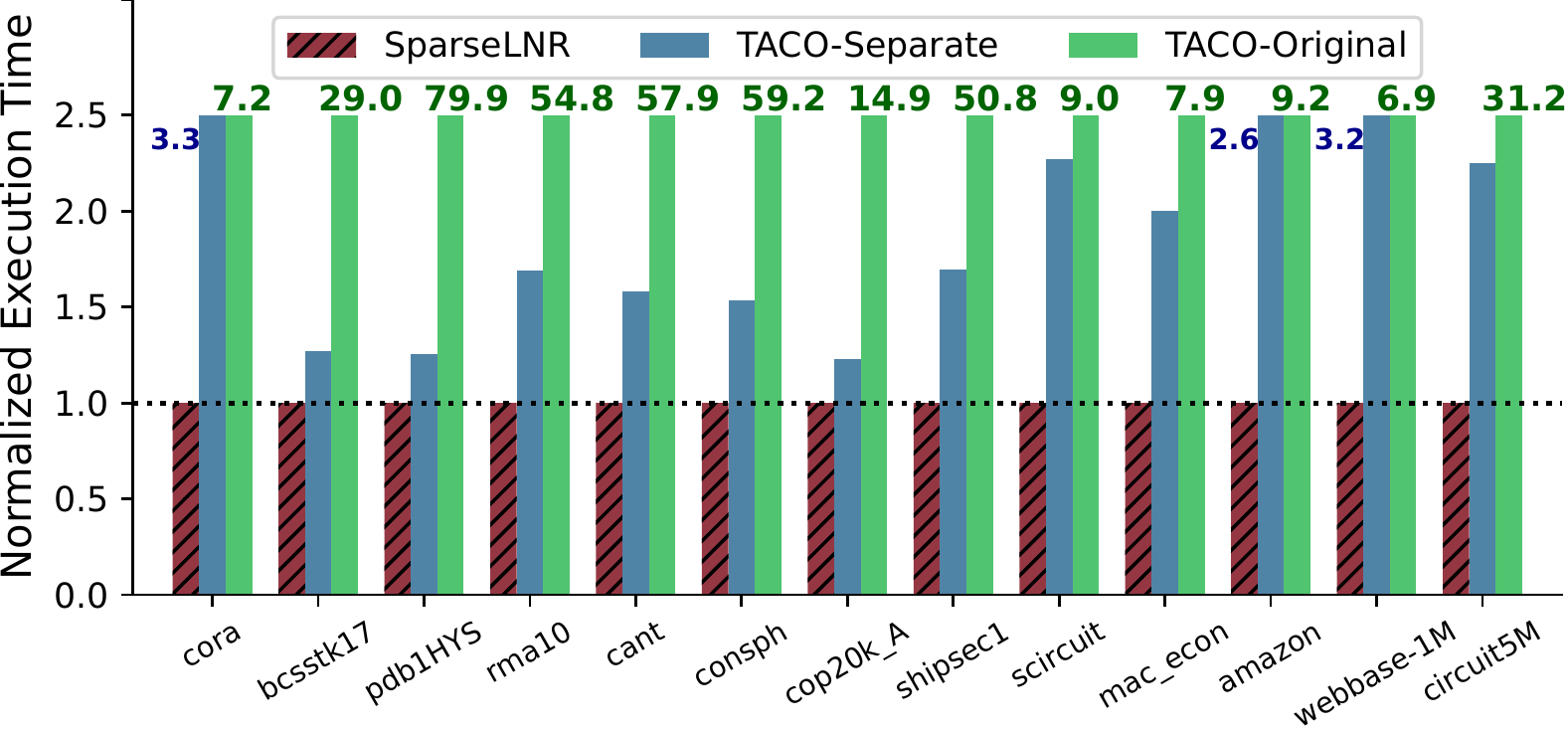}
        \vspace{-4pt}
        \caption{Single-threaded <SpMM, GeMM>}
        \vspace{-10pt}
        \label{fig:spmm-gemm-single-thread}
        \Description[sddmm-spmm original iteration graph]{<long description>}
    \end{subfigure}%
    \hspace{2em}%
    \begin{subfigure}[t]{0.43\textwidth}
        \centering
        \includegraphics[width=0.96\linewidth]{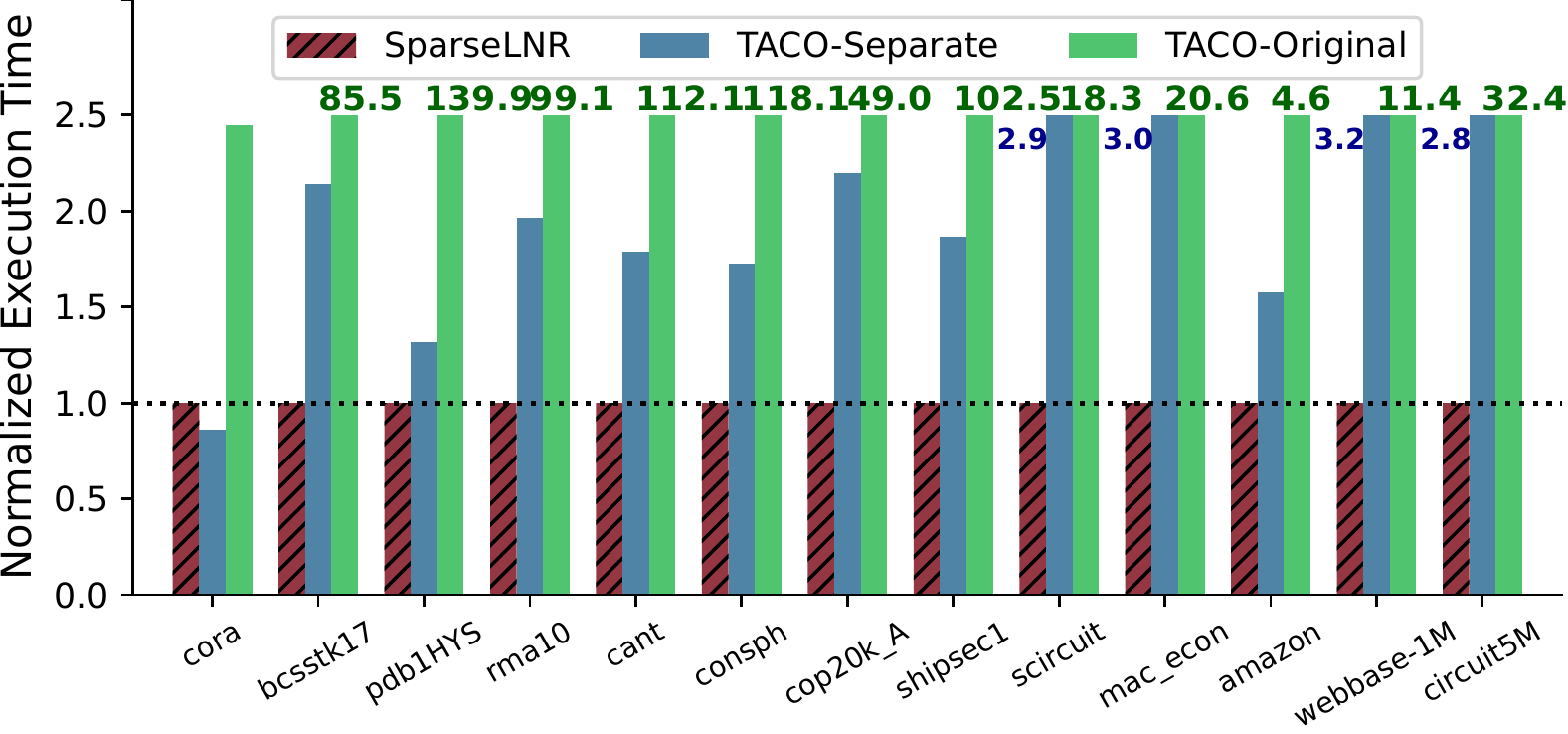}
        \vspace{-4pt}
        \caption{Multi-threaded <SpMM, GeMM>}
        \vspace{-10pt}
        \label{fig:spmm-gemm-multi-thread}
        \Description[sddmm-spmm original iteration graph]{<long description>}
    \end{subfigure}%
    \vskip\baselineskip
    \begin{subfigure}[t]{0.43\textwidth}
        \centering
        \includegraphics[width=0.96\linewidth]{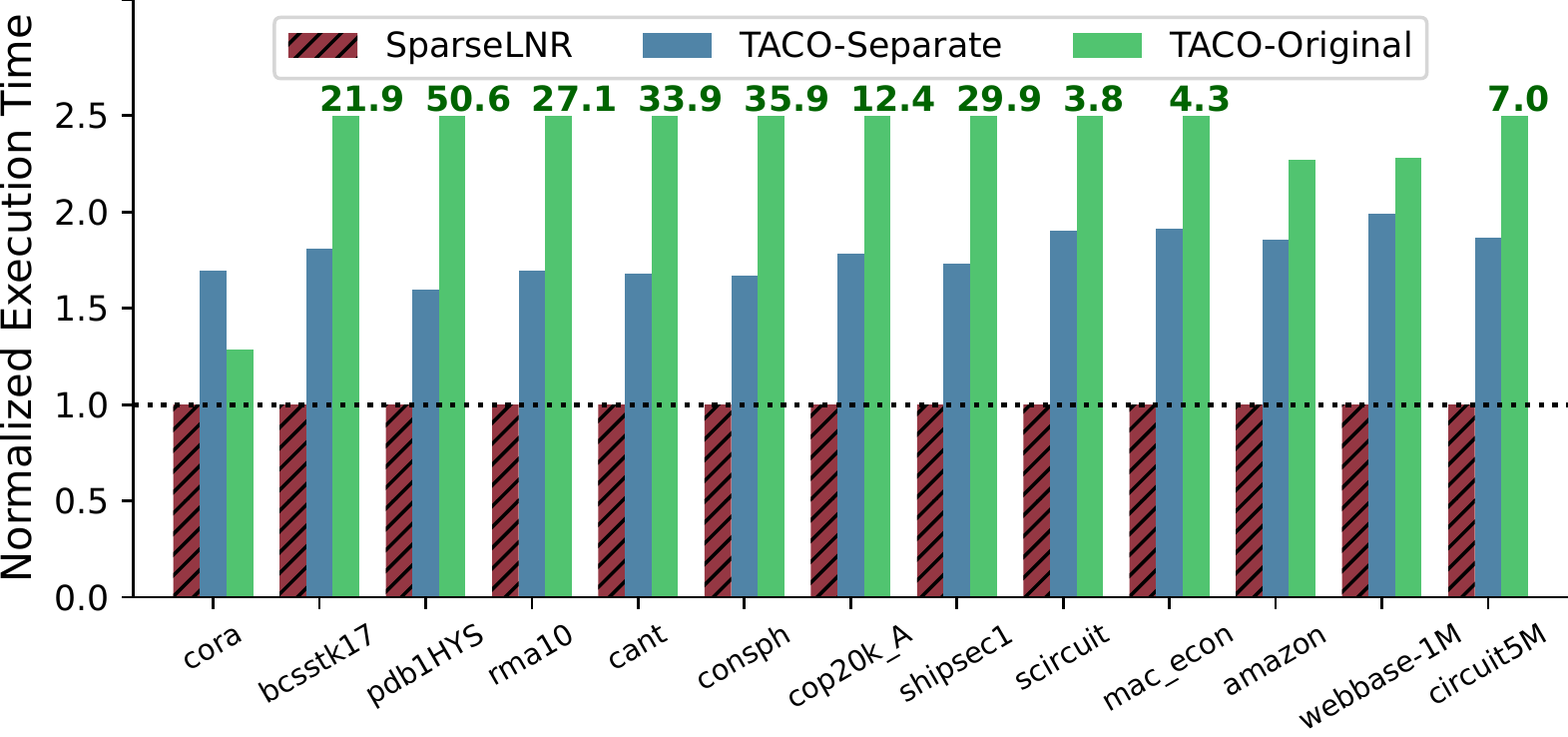}
        \vspace{-4pt}
        \caption{Single-threaded <SpMMH, GeMM>}
        \vspace{-10pt}
        \label{fig:spgemmh-gemm-single-thread}
        \Description[sddmm-spmm original iteration graph]{<long description>}
    \end{subfigure}%
    \hspace{2em}%
    \begin{subfigure}[t]{0.43\textwidth}
        \centering
        \includegraphics[width=0.96\linewidth]{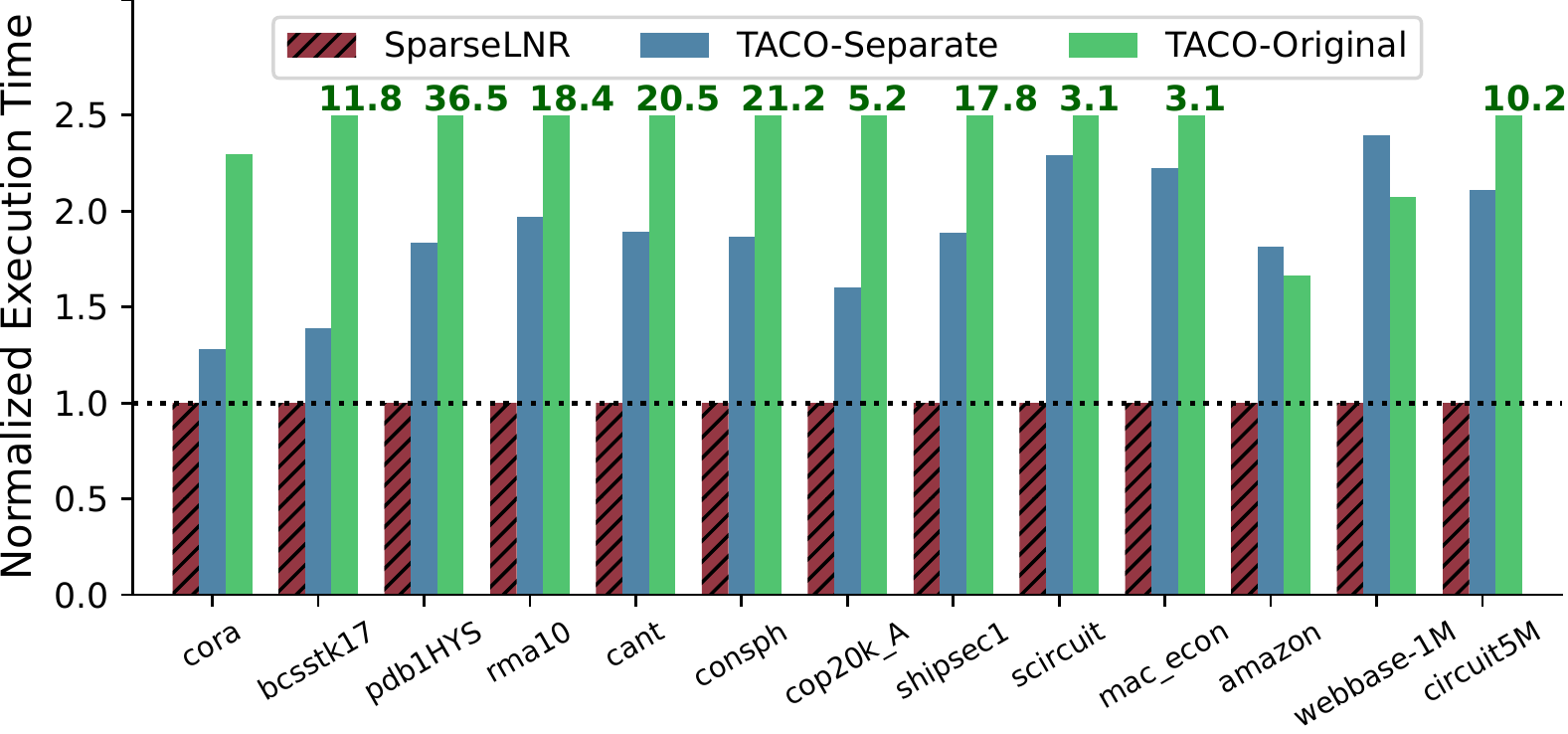}
        \vspace{-4pt}
        \caption{Multi-threaded <SpMMH, GeMM>}
        \vspace{-10pt}
        \label{fig:spgemmh-gemm-multi-thread}
        \Description[sddmm-spmm original iteration graph]{<long description>}
    \end{subfigure}%
    \vskip\baselineskip
    \begin{subfigure}[t]{0.43\textwidth}
        \centering
        \includegraphics[width=0.96\linewidth]{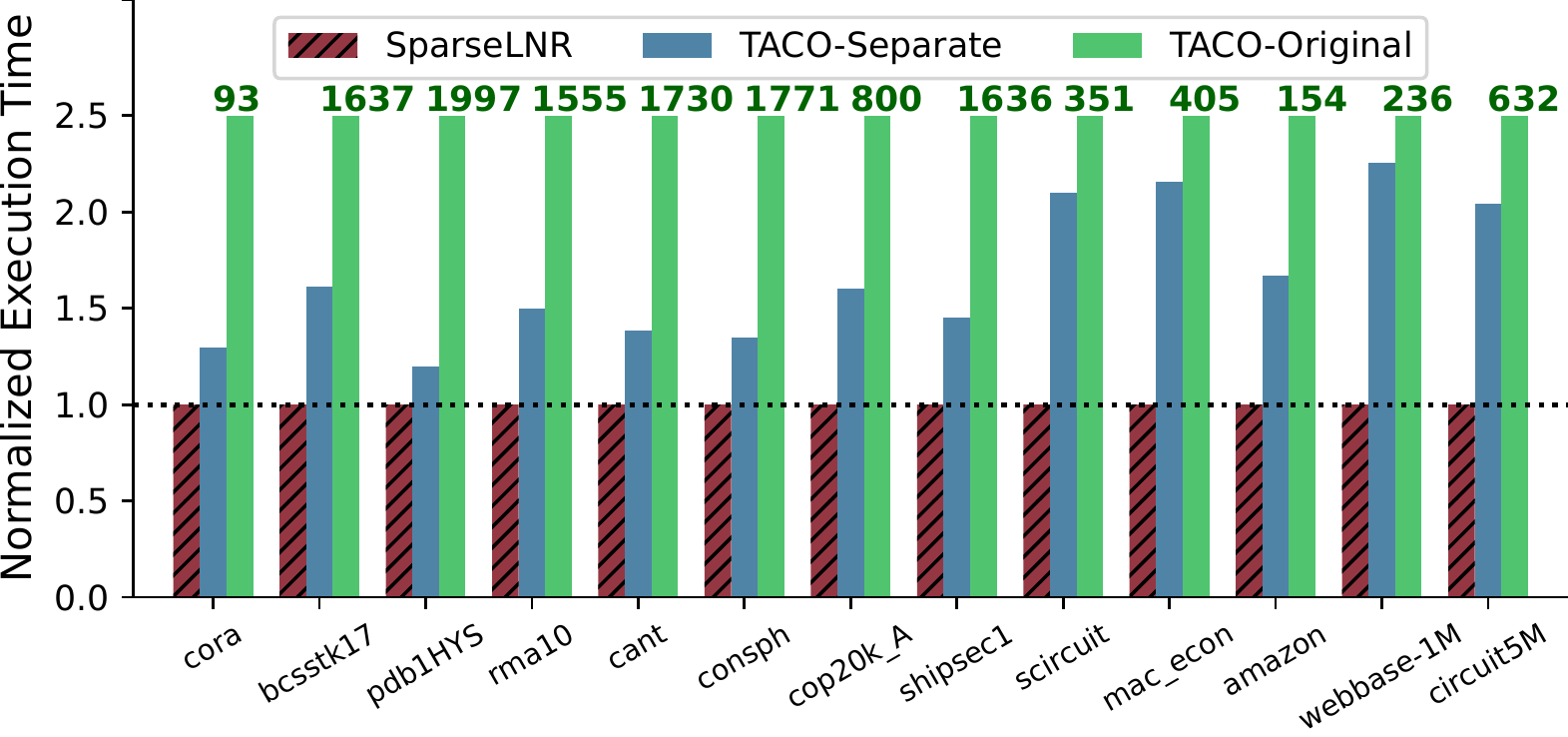}
        \vspace{-4pt}
        \caption{Single-threaded <SDDMM, SpMM, GeMM>}
        \vspace{-10pt}
        \label{fig:sddmm-spmm-gemm-single-thread}
        \Description[sddmm-spmm original iteration graph]{<long description>}
    \end{subfigure}%
    \hspace{2em}%
    \begin{subfigure}[t]{0.43\textwidth}
        \centering
        \includegraphics[width=0.96\linewidth]{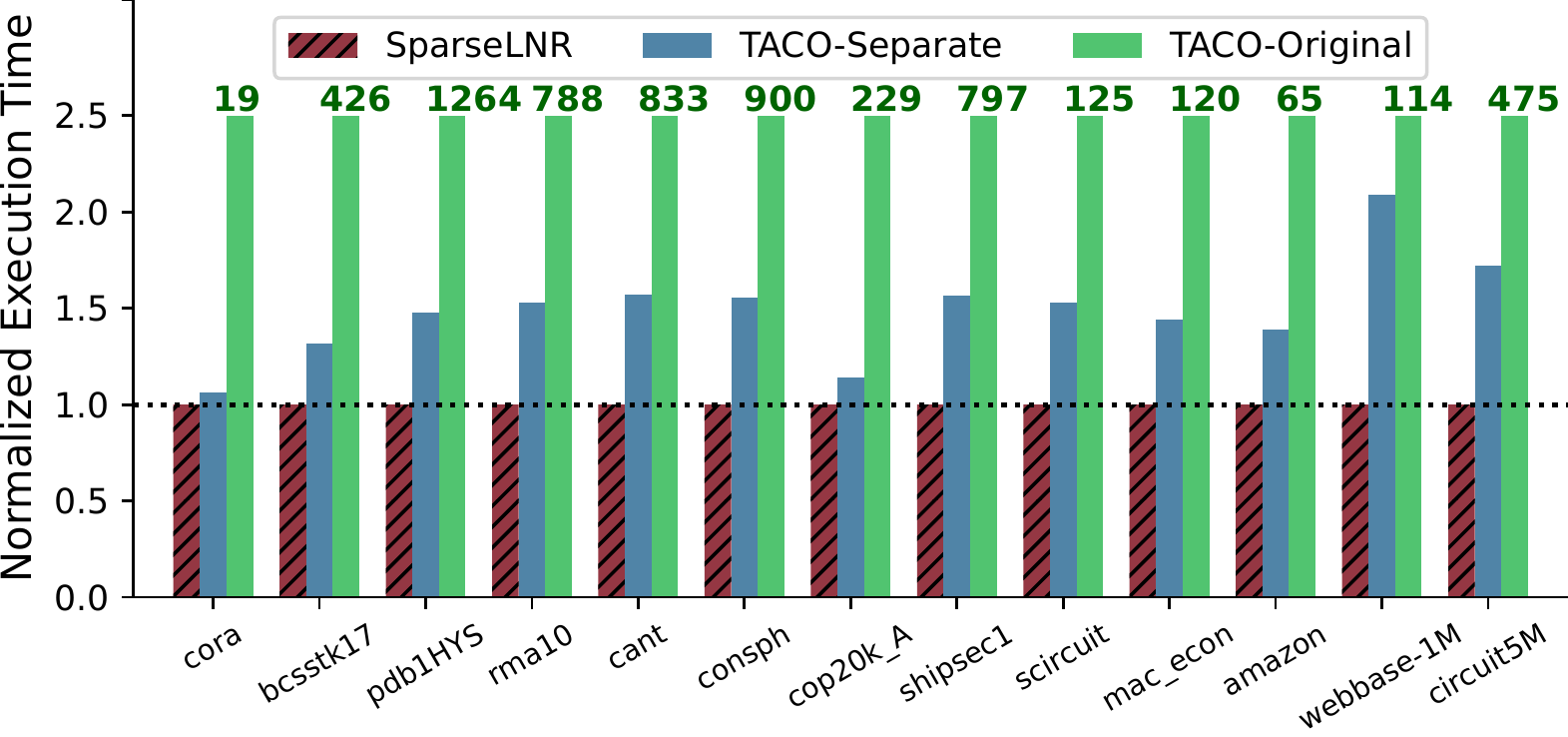}
        \vspace{-4pt}
        \caption{Multi-threaded <SDDMM, SpMM, GeMM>}
        \vspace{-10pt}
        \label{fig:sddmm-spmm-gemm-multi-thread}
        \Description[sddmm-spmm original iteration graph]{<long description>}
    \end{subfigure}%
    \caption{Performance Comparison with TACO for benchmarks with 2-D matrices.}
    \label{fig:evaluation-of-matrices}
    \vspace{-1.5em}
\end{figure*}

\begin{figure*}[ht]
    \vspace{-1.0em}
    \centering
    \begin{subfigure}[t]{0.40\textwidth}
        \centering
        \includegraphics[width=0.96\linewidth]{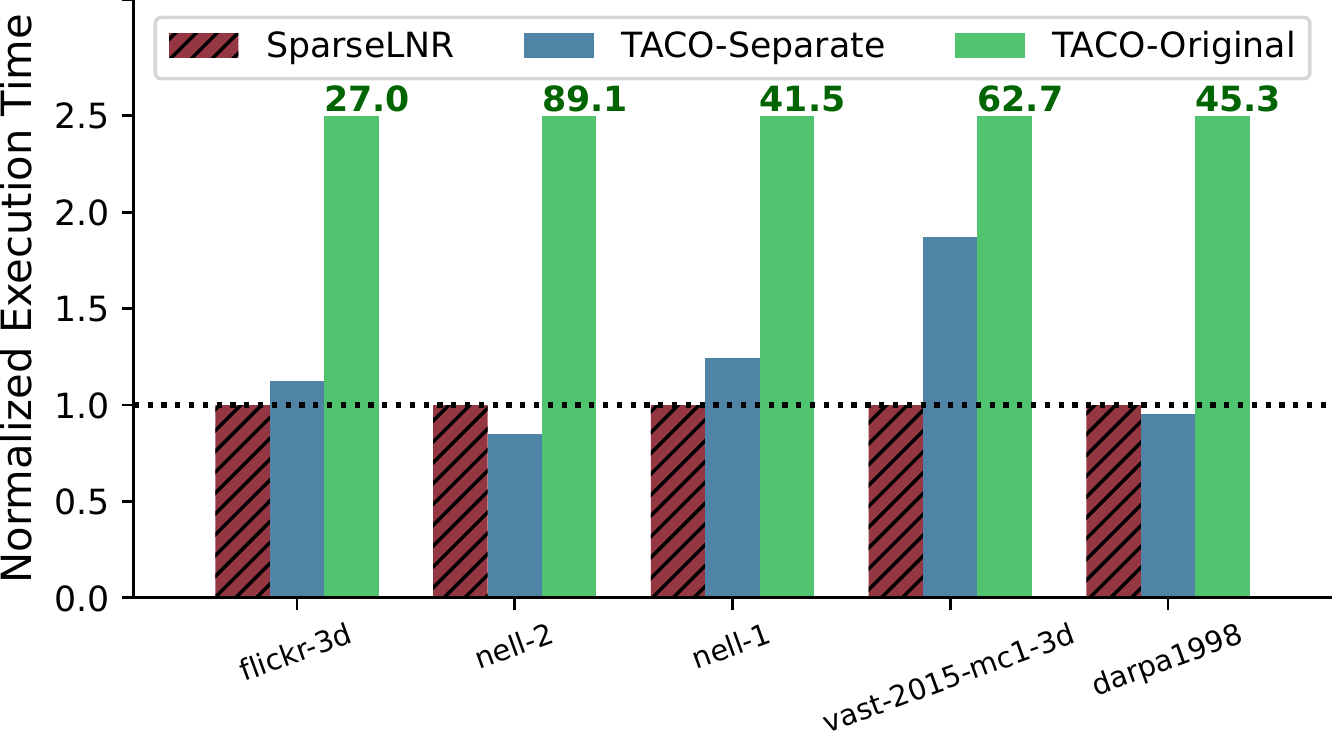}
        \vspace{-4pt}
        \caption{Single-threaded <MTTKRP, GeMM>}
        \vspace{-10pt}
        \label{fig:mttkrp-gemm-single-thread}
        \Description[sddmm-spmm original iteration graph]{<long description>}
    \end{subfigure}%
    \hspace{2em}%
    \begin{subfigure}[t]{0.40\textwidth}
        \centering
        \includegraphics[width=0.96\linewidth]{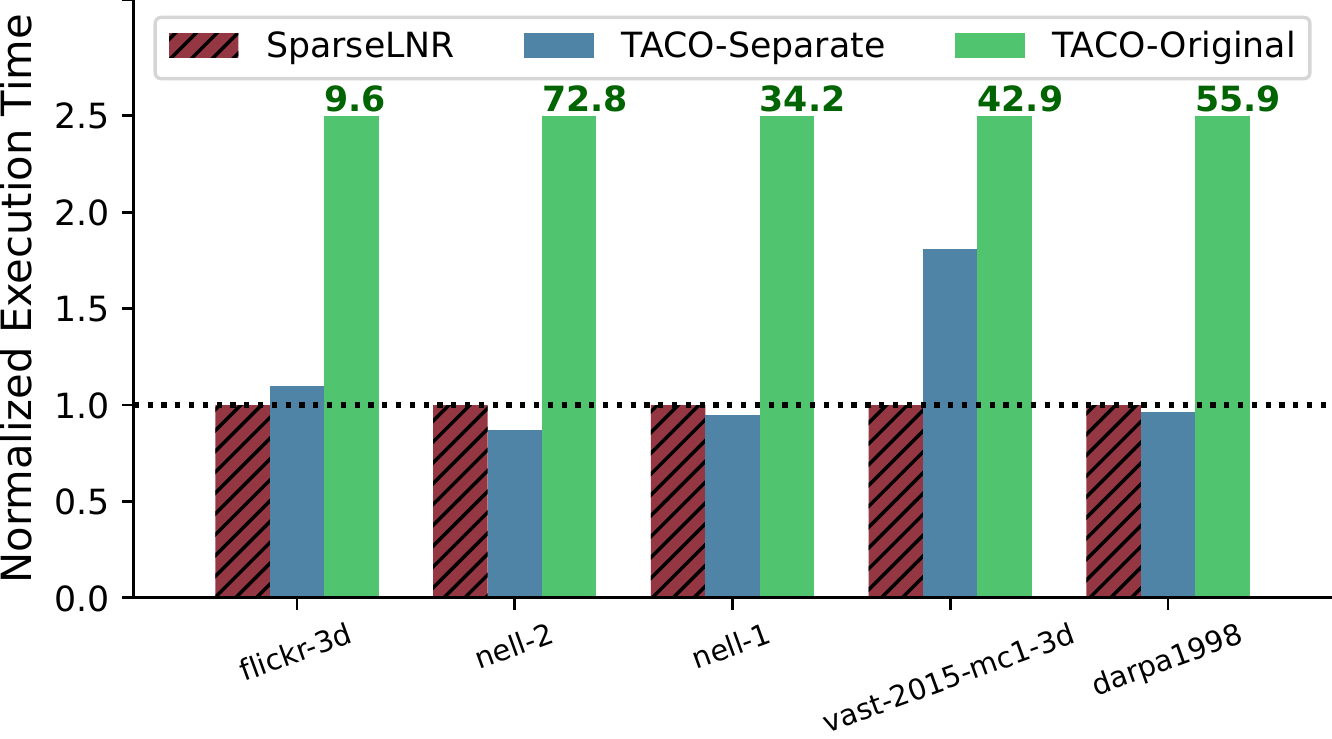}
        \vspace{-4pt}
        \caption{Multi-threaded <MTTKRP, GEMM>}
        \vspace{-10pt}
        \label{fig:mttkrp-gemm-multi-thread}
        \Description[MTTKRP, GEMM multi-threads performance graph]{<long description>}
    \end{subfigure}%
    \vskip\baselineskip
    \begin{subfigure}[t]{0.40\textwidth}
        \centering
        \includegraphics[width=0.96\linewidth]{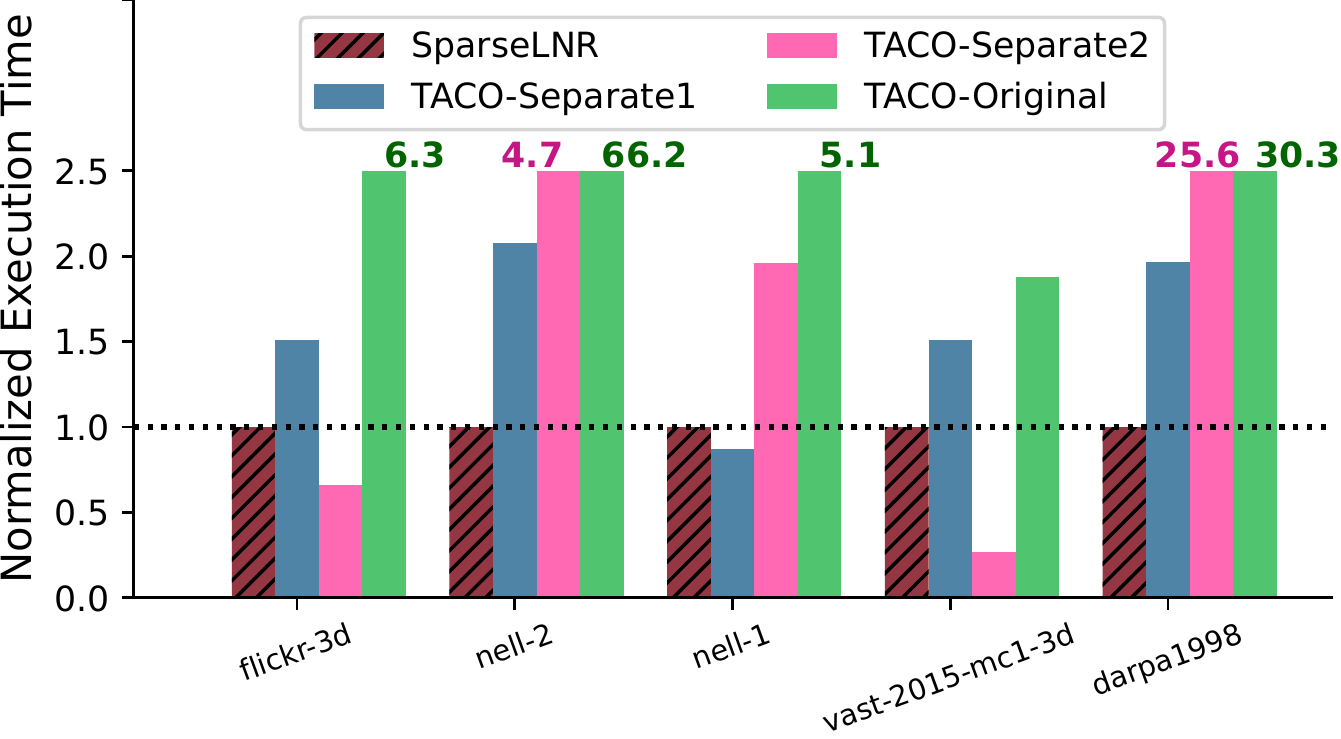}
        \vspace{-4pt}
        \caption{Single-threaded <SpTTM, SpTTM>}
        \vspace{-10pt}
        \label{fig:ttm-ttm-single-thread}
        \Description[sddmm-spmm original iteration graph]{<long description>}
    \end{subfigure}%
    \hspace{2em}%
    \begin{subfigure}[t]{0.40\textwidth}
        \centering%
        \includegraphics[width=0.96\linewidth]{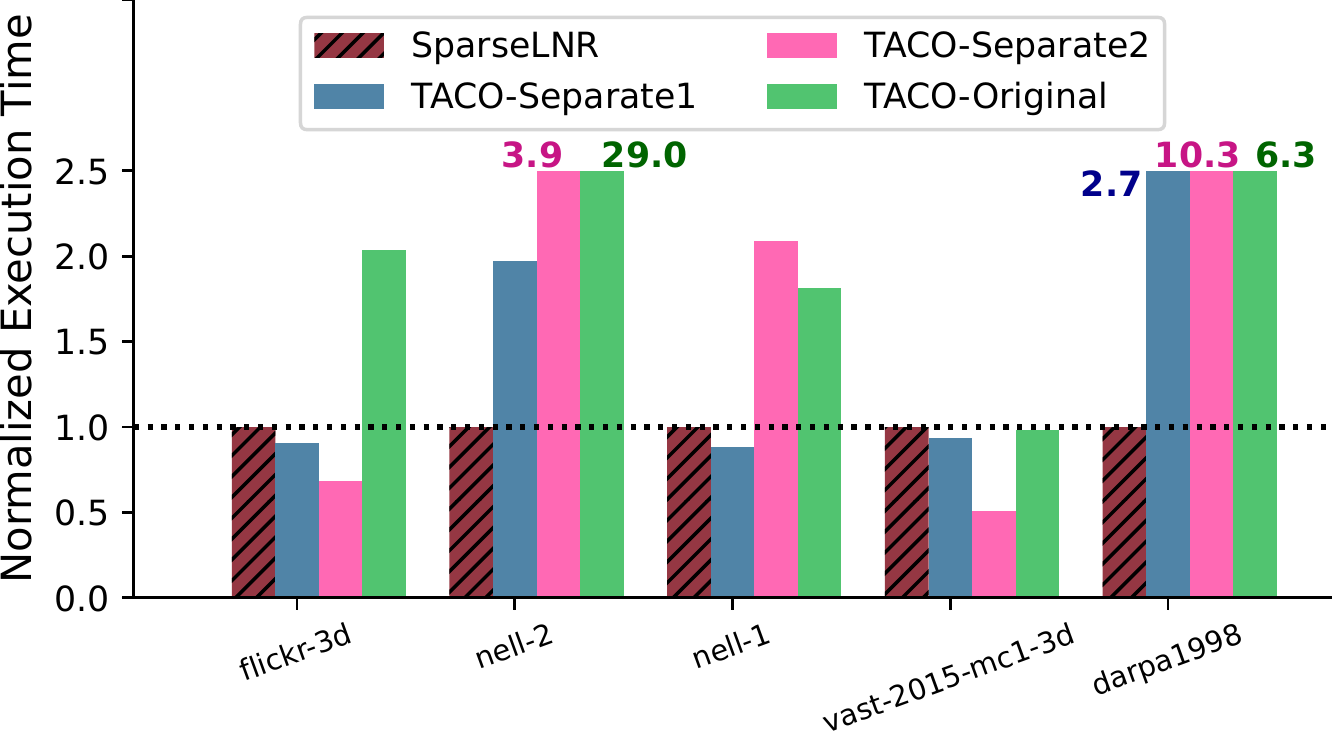}
        \vspace{-4pt}
        \caption{Multi-threaded <SpTTM, SpTTM>}
        \vspace{-10pt}
        \label{fig:ttm-ttm-multi-thread}
        \Description[SpTTM, SpTTM multi-threads performance graph]{<long description>}
    \end{subfigure}%
    \caption{Performance Comparison with TACO for benchmarks with 3-D tensors.}
    \label{fig:evaluation-of-matrices2}
    \vspace{-1.0em}
\end{figure*}

\subsection{Experimental Setup} \label{setup}
All experiments run on a single socket 64-Core AMD Ryzen Threadripper 3990X at 2.2 GHz, with 32KB L1 data cache, 512KB shared L2 cache, and 16MB shared L3 cache. 
We compile the code using GCC 7.5.0 with \code{-O3 -ffast-math}. We use \code{-fopenmp} for parallel versions with {\em OpenMP version 4.5}. 
All parallel versions use 64 threads which is the number of physical cores available in the machine.

\begin{table}[!ht]
    \begin{center}
    \caption{Test tensors used in the evaluation from various matrix and tensor collections mentioned in the Section~\ref{setup}}
    \label{tab:datasets}
    \vspace{-1em}
    \begin{adjustbox}{minipage=\linewidth,scale=0.8}
        \begin{tabular}{l|r|r} 
            \textbf{Tensor} & \textbf{Dimensions} & \textbf{Non-zeros}\\
            \hline%
            cora & { $2.7K \times 2.7K$} & {$5.4K$}\\
            bcsstk17 & { $11K \times 11K$} & {$429K$}\\
            pdb1HYS & { $36K \times 36K$} & {$4.34M$}\\
            rma10 & { $47K \times 47K$} & {$2.37M$}\\
            cant & { $62K \times 62K$} & {$4.01M$}\\
            consph & { $83K \times 83K$} & {$6.01M$}\\
            cop20k\_A & { $12K \times 12K$} & {$2.62M$}\\
            shipsec1 & { $140K \times 140K$} & {$7.81M$}\\
            scircuit & { $171K \times 171K$} & {$959K$}\\
            mac\_econ\_fwd500 & { $207K \times 207K$} & {$1.27M$}\\
            amazon & { $334K \times 334K$} & {$1.85M$}\\
            webbase-1M & { $1.00M \times 1.00M$} & {$3.11M$}\\
            circuit5M & { $5.56M \times 5.56M$} & {$59.52M$}\\
            \hline%
            flickr-3d & { $320K \times 2.82M \times 1.60M$} & { $112.89M$}\\
            nell-2 & { $12K \times 9K \times 288K$} & { $76.88M$}\\
            nell-1 & { $2.9M \times 2.14M \times 25.5M$} & { $143.6M$}\\
            vast-2015-mc1-3d & { $165K \times 11K \times 2$} & { $26.02M$}\\
            darpa1998 & { $22K \times 22K \times 23.7M$} & { $28.42M$}\\
        \end{tabular}
    \end{adjustbox}
    \end{center}
    \vspace{-1.5em}
\end{table}

\noindent
\textbf{Datasets.} \label{datasets} We use sparse tensors from four sources:
SuiteSparse~\cite{suitesparse};
Network Repository~\cite{nr-sigkdd16};
Formidable Repository of Open Sparse Tensors and Tools (FROSTT)~\cite{frosttdataset};
and the 1998 DARPA Intrusion Detection Evaluation~\cite{freebee}.
Dense tensors in kernels are randomly generated. Table~\ref{tab:datasets} gives the details of the sparse tensors.

\subsection{Benchmarks} \label{benchmarks}

\noindent
\textbf{<SDDMM, SpMM>.} SDDMM computation followed by the SpMM operation, $A_{il} = \sum\nolimits {\sparse B_{ij}} \cdot C_{ik} \cdot D_{jk} \cdot E_{jl}$.
This operation is used in graph neural networks~\cite{fusedMM}. We set the inner dimensions $k$ and $l$ to <64, 64>. 
Fusion of SDDMM with SpMM results in a scalar intermediate to share the results between the fused loops as shown in Figure~\ref{fig:SparFF-fused-kernel}.

\noindent
\textbf{<SpMMH, GEMM>.} SpMMH here is pre-multiplying the Hadamard product of two dense matrices by a sparse matrix. 
The combined kernel we evaluate is $ A_{il} = \sum\nolimits {\sparse B_{ik}} \cdot C_{kj} \cdot D_{kj} \cdot E_{jl}  $ with inner dimensions $j$ and $l$ set to <128, 128>. 

\noindent
\textbf{<SpMM, GEMM>.} \label{spmm-gemm-benchmark} SpMM kernel followed by another GEMM kernel, $ A_{il} = \sum\nolimits {\sparse B_{ij}} \cdot C_{jk} \cdot D_{kl} $.
The inner dimensions $k,m$ are set to <128, 64>. 
This calculation is commonly used for updating the hidden state in GNNs~\cite{gnn}. 

\noindent
\textbf{<SDDMM, SpMM, GEMM>.} We combine two of the prior kernels to show the recursive applicability of the algorithm. 
$ A_{im} = \sum\nolimits {\sparse B_{ij}} \cdot C_{ik} \cdot D_{jk} \cdot F_{jl} \cdot W_{lm} $ 
We could relate this execution to performing SDDMM operation to get the attention values along the edges of a graph, multiplying the feature matrix of the graph with a weight matrix to get the new feature set of the graph and then doing a neighbor sum of the graph. 
The inner dimensions $k,l$ and $m$ are set to <64,64,64>.

\noindent
\textbf{<MTTKRP, GEMM>.} Khatri-Rao product (MTTKRP) followed by GEMM operation, $ A_{im} = \sum\nolimits {\sparse B_{ikl}} \cdot C_{lj} \cdot D_{kj} \cdot E_{jm} $.
MTTKRP kernel is used in various sparse computations like signal processing and computer vision~\cite{mttkrp2018}. 
We set the inner dimensions $j$ and $m$ to <32, 64>. 

\noindent
\textbf{<SpTTM, SpTTM>.} Sparse Tensor Times Matrix (SpTTM) operation followed by another SpTTM operation, $ {\sparse A_{ijm}} = \sum\nolimits {\sparse B_{ijk}} \cdot C_{kl} \cdot D_{lm} $.
SpTTM is a computational kernel used in data analytics and data mining applications such as the popular Tucker decomposition~\cite{MA201999}.
The inner dimensions $l$ and $m$ are set to <32,64>. 

\noindent
\textbf{Sparse Formats.}
SpMM, SDDMM kernels use standard compressed sparse row (CSR) format for their sparse matrices whereas SpTTM, MTTKRP kernels use compressed sparse fiber (CSF) format.

For the SDDMM, SpMM, MTTKRP kernels in {\em TACO separate} we use the versions provided in Senanayake \etal~\cite{senanayake:2020:scheduling}.
For the rest of kernels we evaluate multiple schedules and select the best performing one. 
TACO does not generate multi-threaded code when the output tensor is sparse. Prior work has evaluated against single-threaded code in such situations~\cite{kjolstad:2018:workspaces,compiler_in_mlir}. Following the strategy of Senanayeke \etal~\cite{senanayake:2020:scheduling}, we modified the TACO generated code manually to add multithreading

In general, the speedups we see compared to the TACO original comes from the reduction in asymptotic complexity while the speedups we see compared to the TACO separate comes from the reduction in cache reuse distances by removing large tensors used to store intermediate results.

We see speedups for our approach of 0.90--1.23x compared TACO Separate and 3.31--16.05 compared to TACO Original in <SDDMM, SpMM> kernel's multi-threaded execution.
For single-threaded execution, we get 0.91--1.50x compared to TACO separate and 10.75--33.39x compared to TACO original.
In multithreaded execution the fused kernel performs better on circuit5M, shipsec1, consph, pdf1HYS, and cant, the matrices with most non-zeros, from the tested datasets~\ref{tab:datasets} compared against the separate execution.

We observe speedups of 1.60--1.99x against TACO separate and 1.29--50.55x against TACO original in single-threaded execution for <SpMMH, GEMM> kernel.
For the same kernel in multi-threaded execution, we observed speedups of 1.28--2.40x against TACO separate and 1.66--36.50x against TACO original.
For the <SpMM, GEMM> kernel in single-threaded execution, speedups of 1.23--3.27x, and 6.91--79.86x are observed for TACO original and TACO separate, respectively.
For the same kernel in multi-threaded execution, we observed speedups of 0.86--3.16x, and 2.44--139x for TACO original and TACO separate, respectively.

We see substantial speedups in <SDDMM, SpMM, GEMM> due to the kernel presenting two opportunities for fusion: <SDDMM, SpMM> and <SpMM, GEMM>.
We see speedups ranging from 1.20--2.26x for single-threaded execution against TACO separate, 93--1997x over single-threaded TACO original, 1.06--2.09x for multithreaded TACO separate and 19--1263x for multithreaded TACO original.

We see that our approach under-performed in datasets nell-2 and darpa1998 against the TACO separate for the <MTTKRP, GEMM> benchmark. 
In these datasets, the first dimensions of the tensors 
are bounded by 12092, and 22476. 
Therefore, the intermediate matrix in the TACO separate execution has sizes that fit within the last level cache. 
Furthermore, executing kernels separately sometimes offer more opportunities to optimize smaller kernels individually. 
Due to these reasons, there may be datasets that the separate kernel execution performs better than the fused version. 
But we see considerable speedups versus TACO Separate when the intermediate tensors are large.

\begin{figure}[!t]
    \centering
    \rulesep%
    \begin{subfigure}[t]{0.22\linewidth}
        \centering
        \includegraphics[width=1\linewidth]{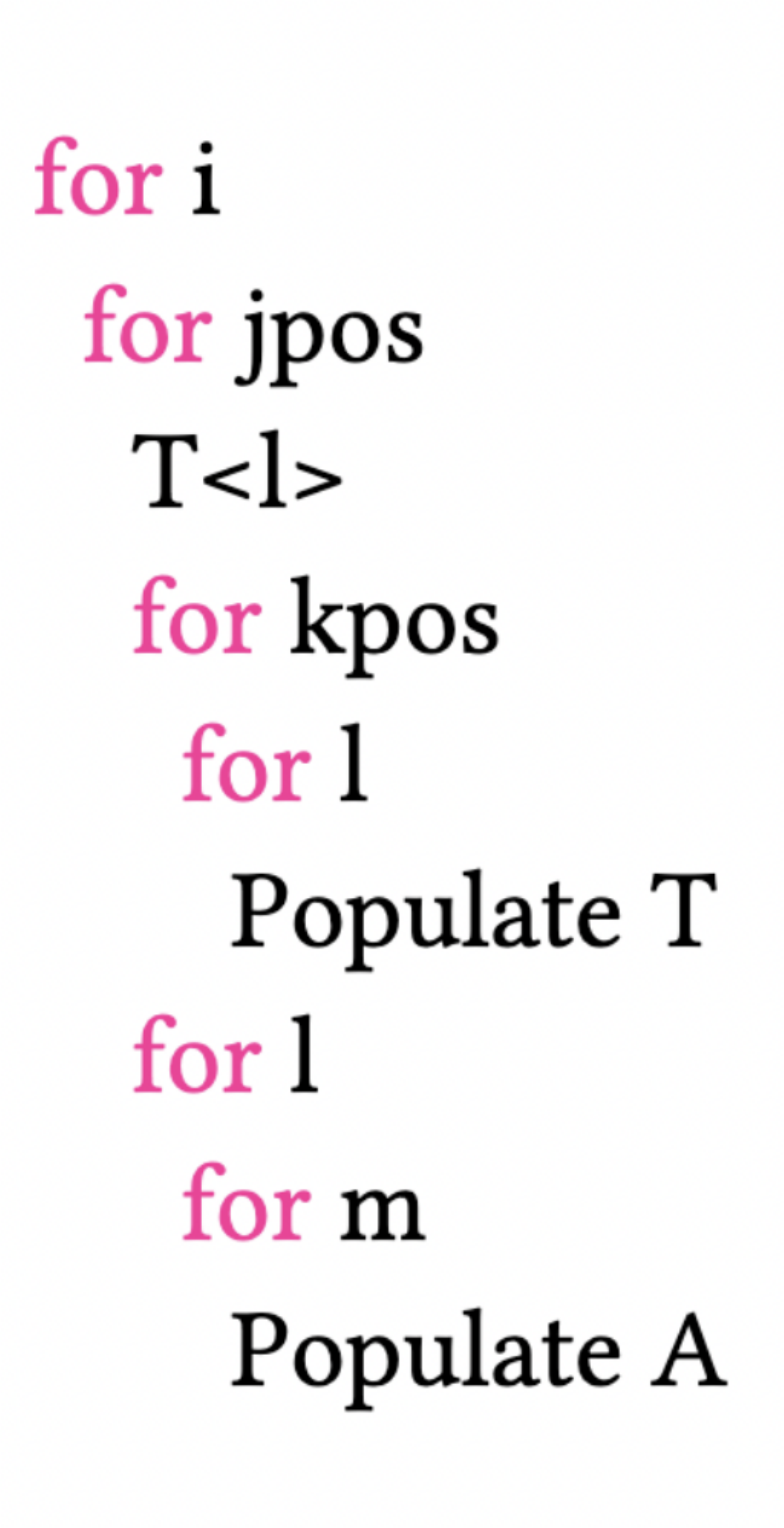}
        \vspace{-15pt}
        \caption{\ }
        \vspace{-10pt}
        \label{fig:fusedttm}
        \Description[Branched iteration graph for the SDDMM, SpMM operation]{<long description>}
    \end{subfigure}%
    \rulesep%
    \begin{subfigure}[t]{0.22\linewidth}
        \centering
        \includegraphics[width=1\linewidth]{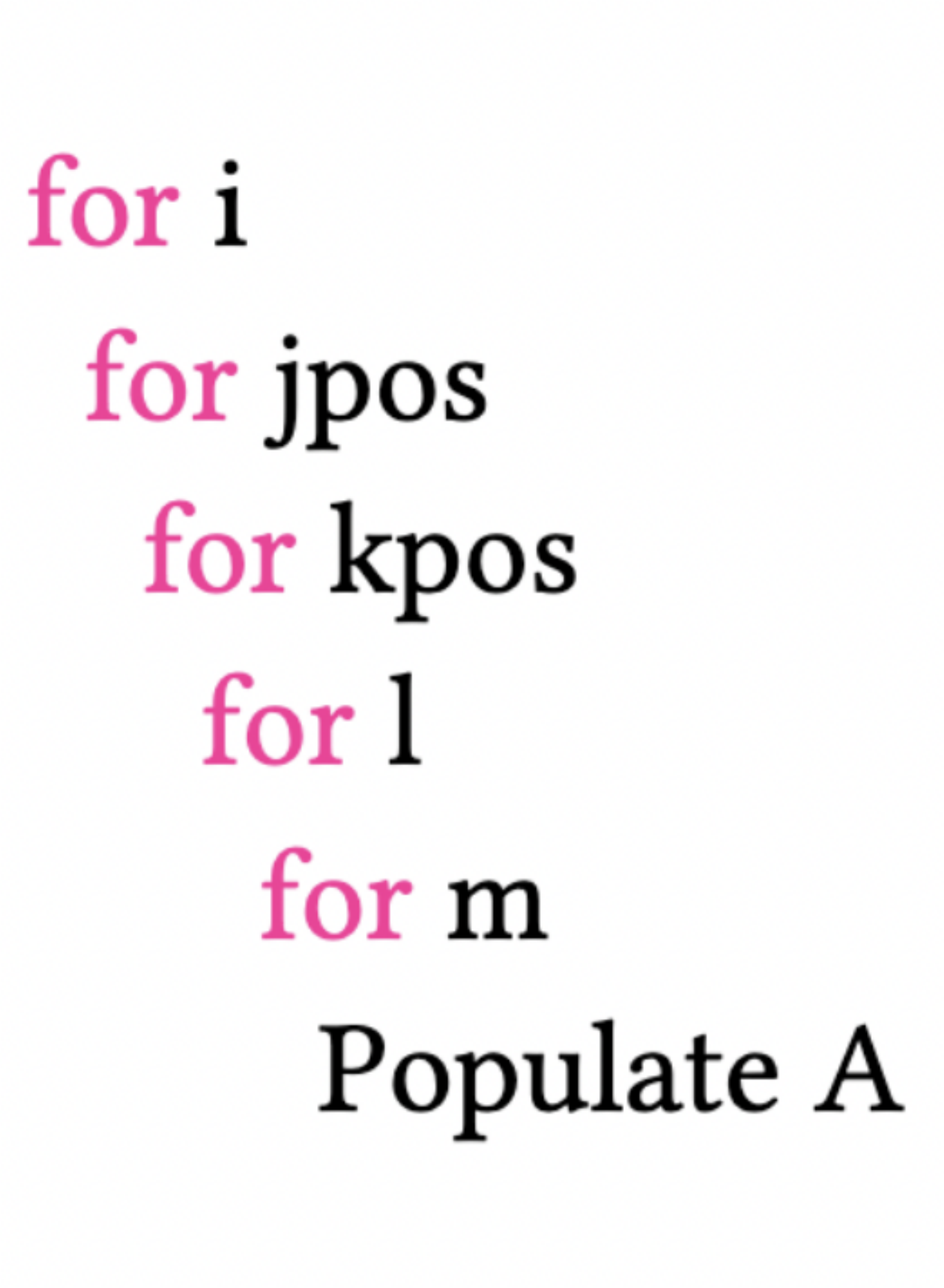}
        \vspace{-15pt}
        \caption{\ }
        \vspace{-10pt}
        \label{fig:originalttm}
        \Description[Branched iteration graph for the SDDMM, SpMM operation]{<long description>}
    \end{subfigure}%
    \rulesep%
    \begin{subfigure}[t]{0.22\linewidth}
        \centering
        \includegraphics[width=1\linewidth]{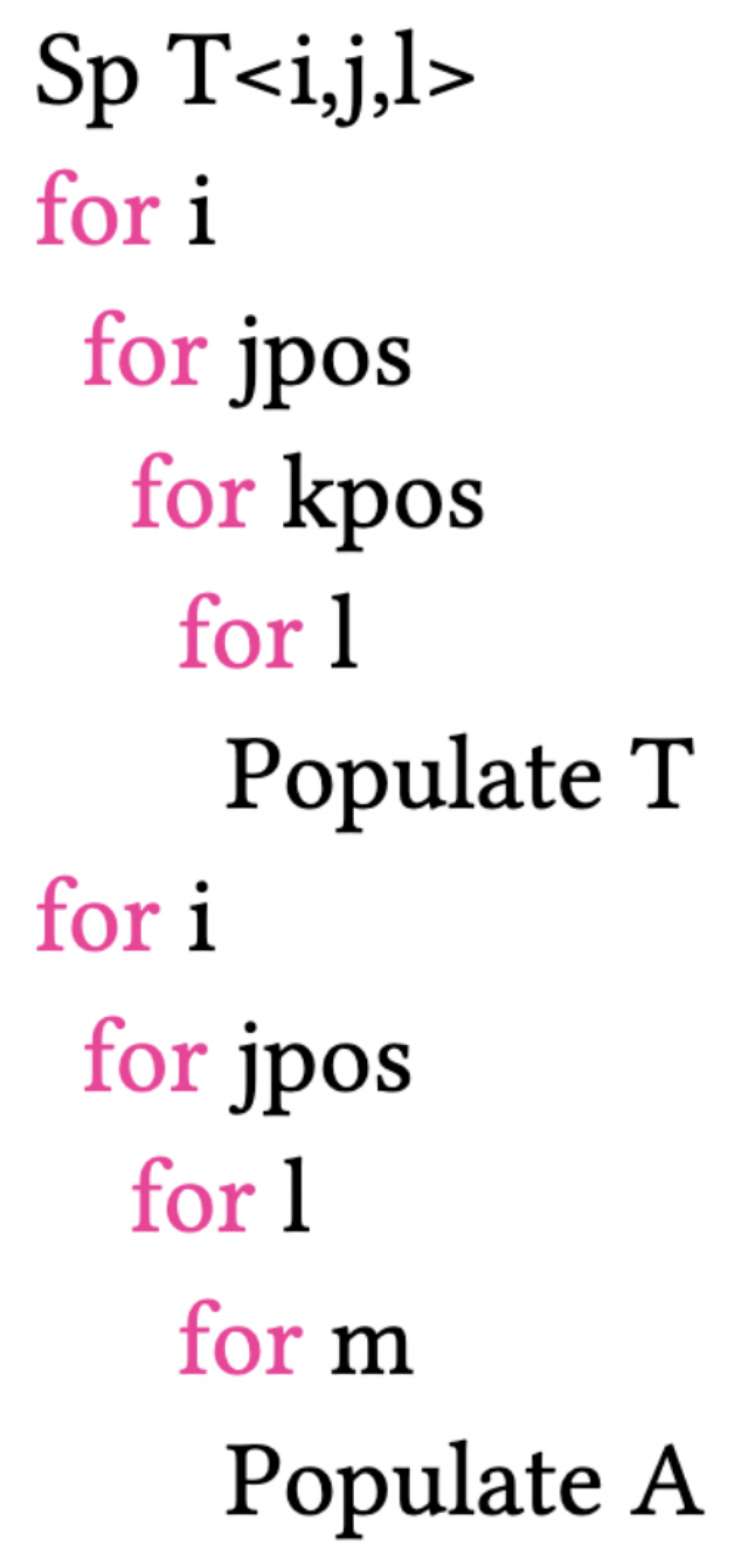}
        \vspace{-15pt}
        \caption{\ }
        \vspace{-10pt}
        \label{fig:separate1ttm}
        \Description[Branched iteration graph for the SDDMM, SpMM operation]{<long description>}
    \end{subfigure}%
    \rulesep%
    \begin{subfigure}[t]{0.22\linewidth}
        \centering
        \includegraphics[width=1\linewidth]{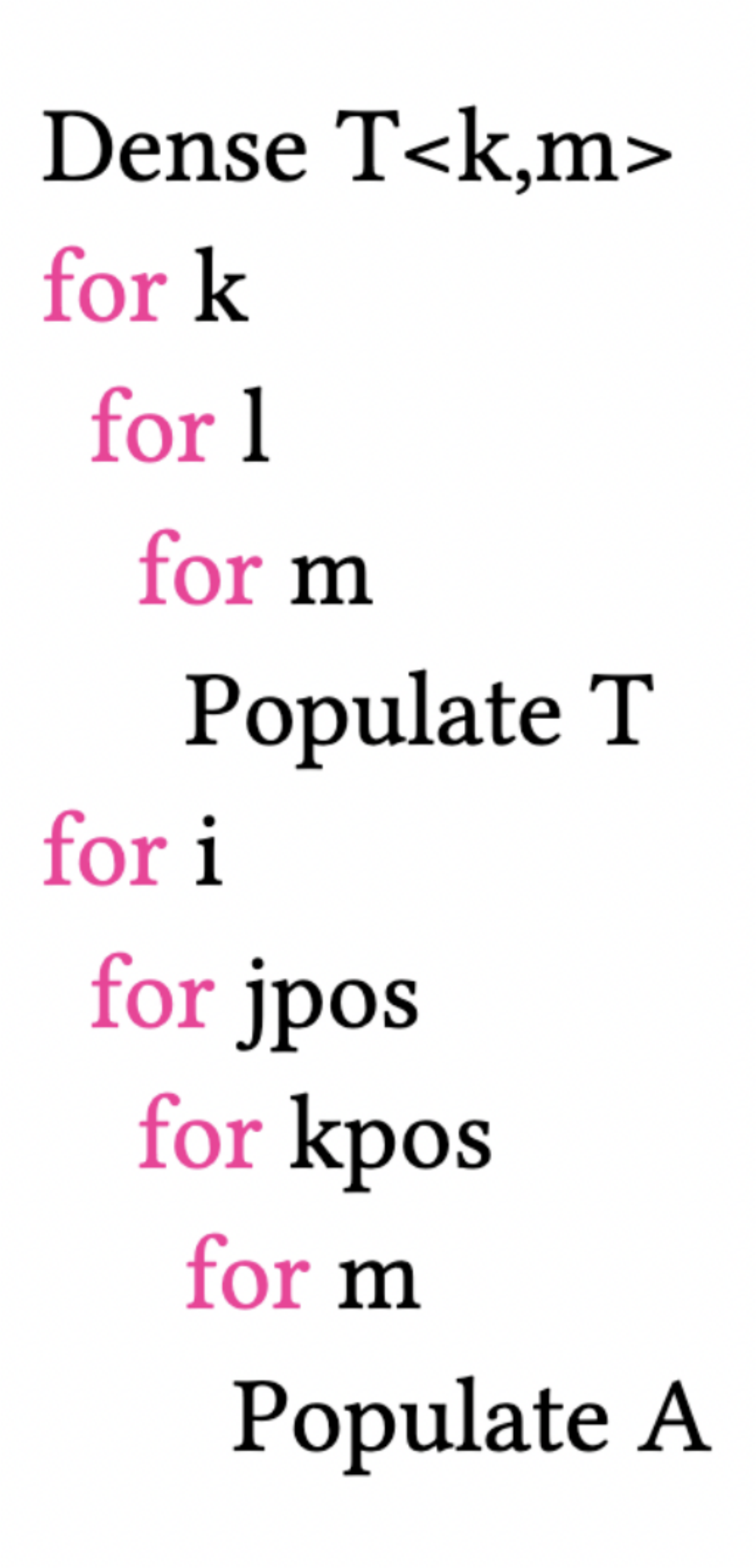}
        \vspace{-15pt}
        \caption{\ }
        \vspace{-10pt}
        \label{fig:separate2ttm}
        \Description[Branched iteration graph for the SDDMM, SpMM operation]{<long description>}
    \end{subfigure}%
    \rulesep%
    \caption{Basic loop structure of different schedules for <SpTTM, SpTTM> kernel.}
    \label{fig:ttmloopstructure}
    \vspace{-2.0em}
\end{figure}

For <SpTTM, SpTTM>, there are two different schedules due to different associativity choices. We see varying results based on which choice is made, so we report both of them as {\em separate1} and {\em separate2} (see Figures~\ref{fig:ttm-ttm-single-thread} 
and~\ref{fig:ttm-ttm-multi-thread}). 
Figure~\ref{fig:ttmloopstructure} shows the basic loop structure for different versions of <SpTTM, SpTTM>.
The fused version has asymptotic complexity of $O(nnz(B_{IJK})L+nnz(B_{IJ})LM)$. The TACO original version $ {\sparse A_{ijm}} = \sum\nolimits {\sparse B_{ijk}} \cdot C_{kl} \cdot D_{lm} $ 
has  asymptotic complexity of $O(nnz(B_{IJK})LM)$. But separate1 
($ {\sparse T_{ijl}} = \sum\nolimits {\sparse B_{ijk}} \cdot C_{kl} $ followed by 
$ {\sparse A_{ijl}} = \sum\nolimits {\sparse T_{ijl}} \cdot C_{kl} $) and 
separate2 ($ T_{km} = \sum\nolimits C_{kl} \cdot D_{lm} $ followed by 
$ {\sparse A_{ijm}} = \sum\nolimits {\sparse B_{ijk}} \cdot T_{km} $) versions have complexities of 
$O(nnz(B_{IJK})L+nnz(B_{IJL})M)$ and 
$O(KLM+nnz(B_{IJK})M)$, respectively. 

When $k$ is small (for instance, dimension $k$ of the dataset {\em vast2015-mc1-3d} in Table~\ref{tab:datasets} is bounded by 2), the asymptotic complexity of our approach is comparable to TACO's baseline approach, and the size of the working set is small. 
Therefore, there are no cache misses for the TACO separate approach; the re-associated schedule hence has the best performance. 
In the other datasets, where $k$ is large (for instance, the dimension $k$ of the dataset {\em darpa1998} in Table~\ref{tab:datasets} is bounded by 28.42M), these effects vanish, 
and our approach is considerably faster than any competing version. We note that \system's representation could support re-association scheduling directives, that would allow it to use the better schedules of TACO separate, but we leave that for future work.

\begin{figure}[!ht]
    \centering
    \begin{subfigure}[t]{0.80\columnwidth}
        \centering
        \includegraphics[width=1\linewidth]{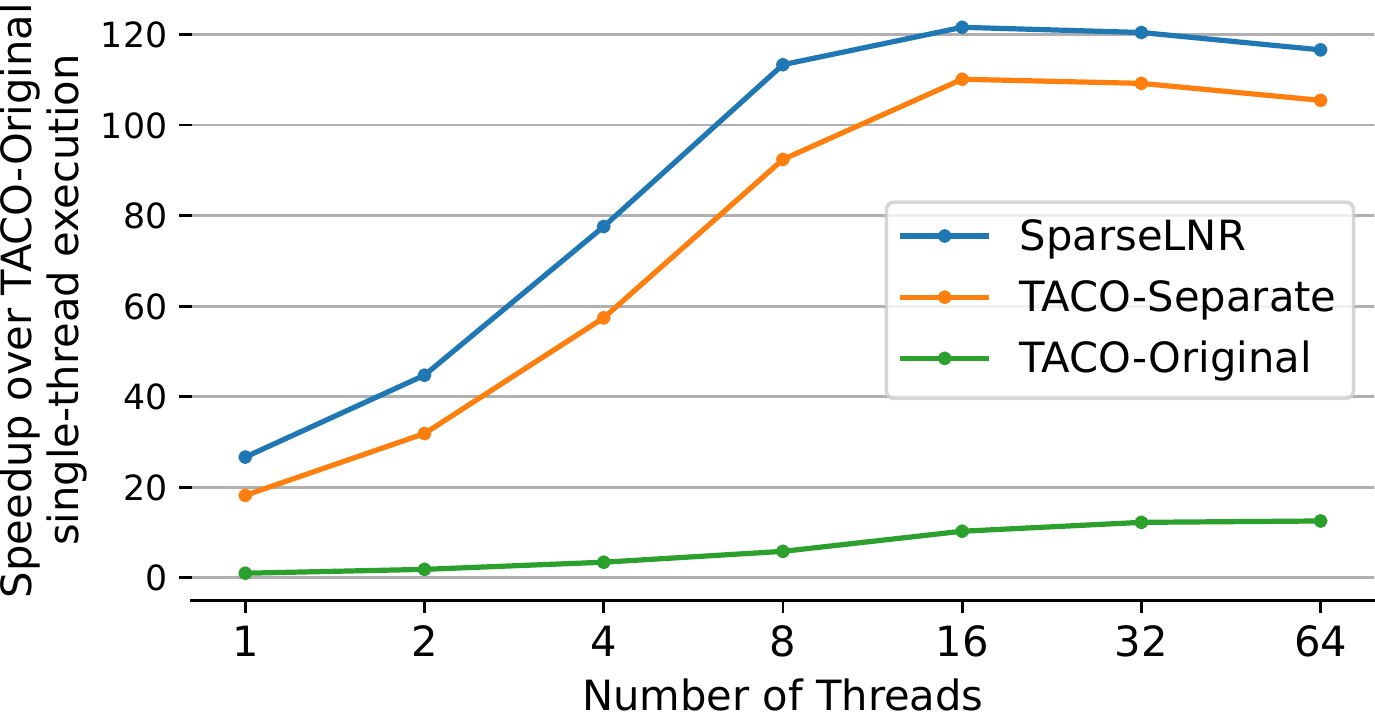}
        \vspace{-15pt}
        \caption{Circuit5M Scaling}
        \vspace{-15pt}
        \label{fig:circuit5M-speedup}
        \Description[sddmm-spmm original iteration graph]{<long description>}
    \end{subfigure}%
    \vskip\baselineskip
    \begin{subfigure}[t]{0.80\columnwidth}
        \centering
        \includegraphics[width=1\linewidth]{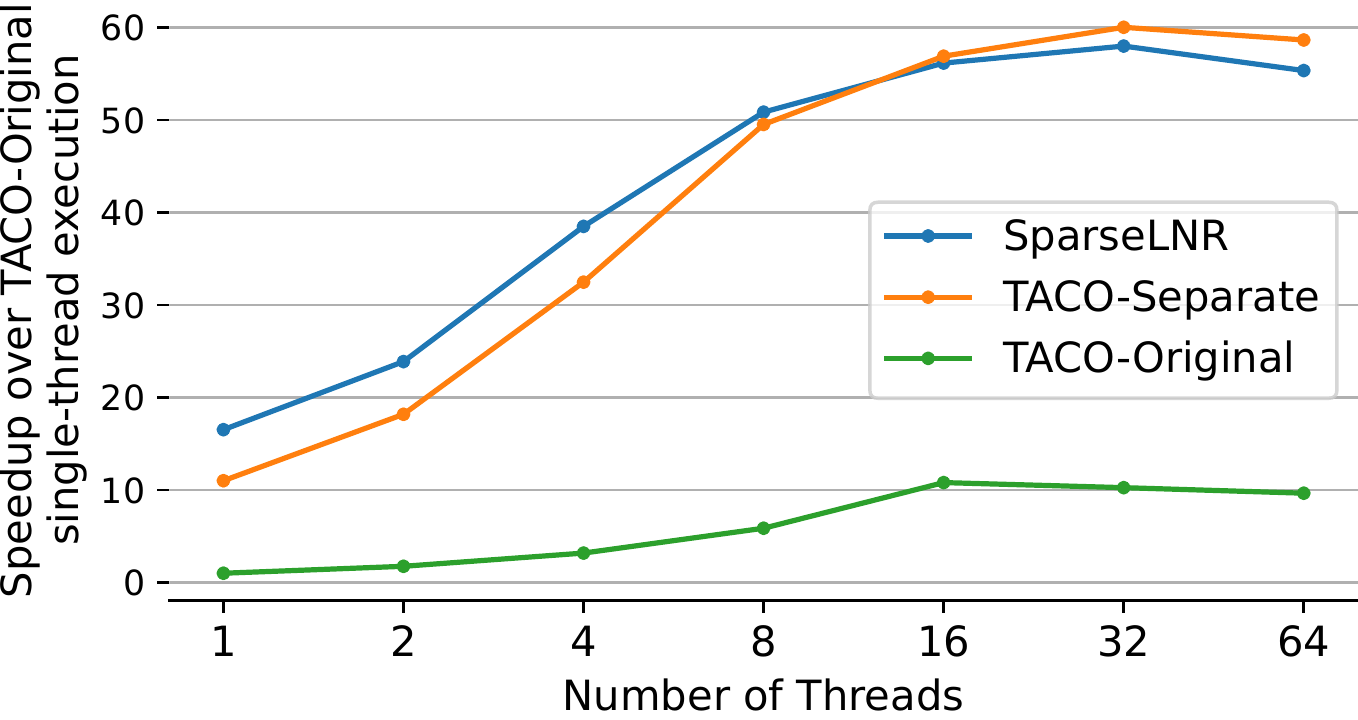}
        \vspace{-15pt}
        \caption{Webbase Scaling}
        \vspace{-12pt}
        \label{fig:webbase-speedup}
        \Description[sddmm-spmm original iteration graph]{<long description>}
    \end{subfigure}%
    \caption{Speedup change with respect to the number of threads for <SDDMM, SpMM> benchmark.}
    \label{fig:threading-speedup}
    \vspace{-1.9em}
\end{figure}
The scaling results for webbase and circuit5M datasets, on <SDDMM, SpMM> benchmark are shown in Figure~\ref{fig:threading-speedup}; \system delivers comparable scaling to the best alternative approach.
We observed similar scaling for other benchmarks and datasets.

\subsection{Case Study: Performance with different inner dimensions} \label{casestudy1}

\begin{figure}[!ht]
    \centering
    \begin{subfigure}[t]{0.45\columnwidth}
        \includegraphics[width=\linewidth]{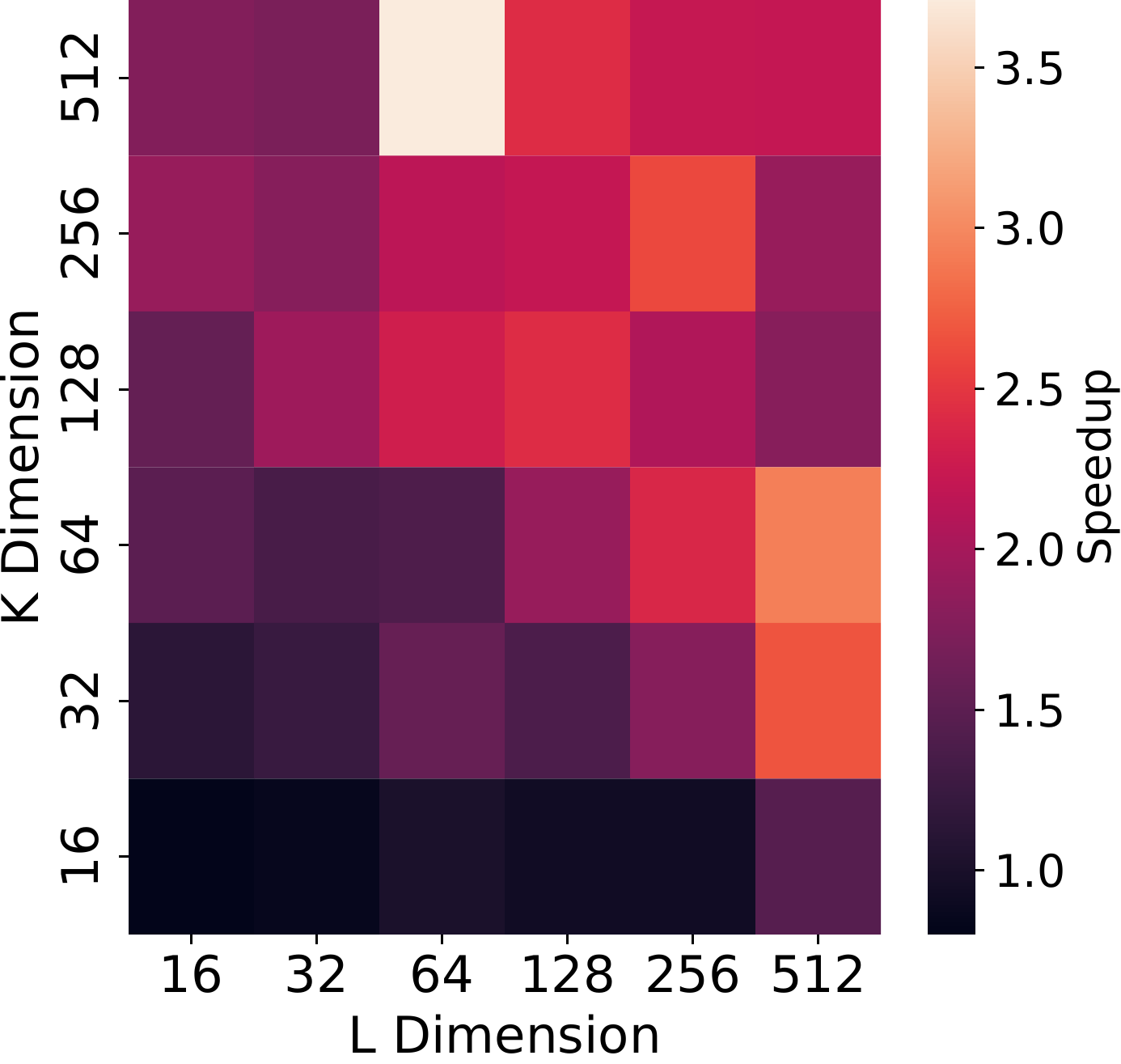}
        \caption{TACO-Sep. vs. \system}
        \vspace{-10pt}
        \label{fig:heatmap-with-taco-separate}
    \end{subfigure}%
    \hspace{1em}%
    \begin{subfigure}[t]{0.45\columnwidth}
        \includegraphics[width=\linewidth]{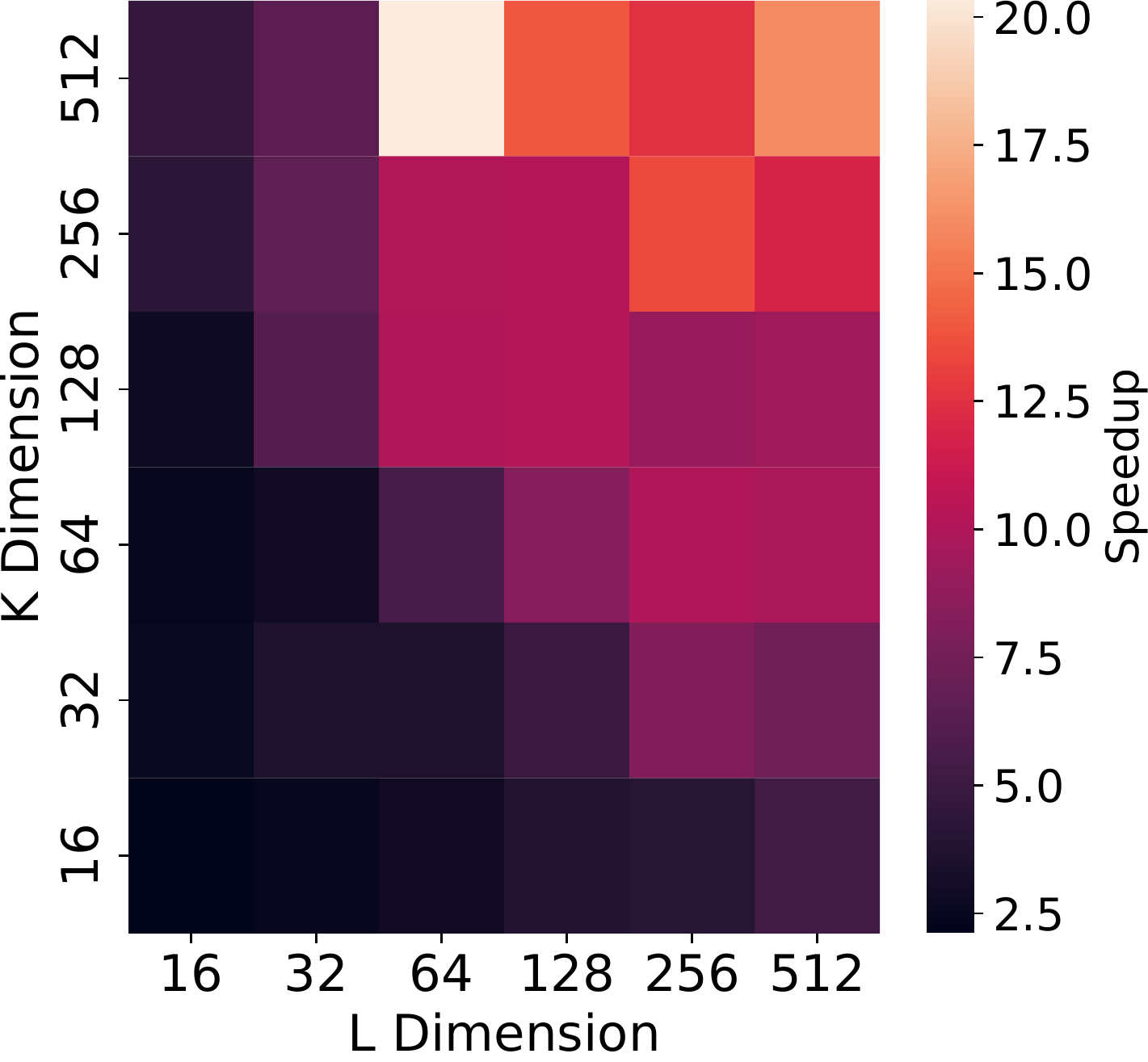}
        \caption{TACO-Orig. vs. \system}
        \vspace{-10pt}
        \label{fig:heatmap-with-taco-original}
    \end{subfigure}%
    \caption{Speedup variation wrt. inner dimension sizes ($k$ and $l$) on the benchmark <SpMM,GEMM>. The Figures (a) and (b) correspond to multithreaded execution of \system against TACO-Separate and TACO-Original, respectively.}
    \label{fig:heatmap-speedup}
    \vspace{-1.5em}
\end{figure}

We chose the inner dimensions of the benchmarks explained in Section~\ref{benchmarks} arbitrarily.
In this case study we consider change of performance wrt. varying inner dimension sizes ($k$ and $l$) in the benchmark <SpMM, GEMM> ($ A_{il} = \sum\nolimits {\sparse B_{ij}} \cdot C_{jk} \cdot D_{kl} $) in Section~\ref{spmm-gemm-benchmark}.
Here, the dimensions $i$ and $j$ are determined by size of the graphs read into the sparse matrix $\sparse B_{ij}$ and cannot be arbitrarily changed.
However, the dimensions $k$ and $l$ can change in size since the matrices $C_{jk}$ and $D_{kl}$ are dense. 
Usually, in GNN literature, these dimensions correspond to feature sizes in hidden layers.

Performing the \emph{loop fusion after distribution} in the benchmark <SpMM, GEMM> results in a temporary vector of the length of the size of dimension $k$, and $k$ and $l$ dimensions completely define the size of the matrix $D_{kl}$. 
When the size of the dimension $l$ is small, $D_{kl}$ can completely fit in higher level caches for the $k$ values considered.
Therefore, for smaller $l$ values, the speedup increases with the size of $k$ (See Figure~\ref{fig:heatmap-speedup}).
Because with increasing $k$, the temporary tensor gets larger and it decreases the performance of TACO-Separate.
We see this behavior for $l$ dimension sizes of 16-128 in Figure~\ref{fig:heatmap-with-taco-separate}.
But with increasing $k$, when the size of the $l$ dimension gets larger, the sizes of $D_{kl}$ and temporary vector get larger.
As a result, they keep getting evicted from the higher level caches in \system. 
Therefore, as shown in Figure~\ref{fig:heatmap-with-taco-separate}, the peak performance in columns 256 and 512 of the dimension $l$ occurs not when $k$ is 512, but when size of $k$ takes values in the range 64--256.
Increasing sizes of $k$ and $l$ results in higher time complexity for TACO-Original. 
Hence, speedup of \system increases with the sizes of $k$ and $l$ in the Figure~\ref{fig:heatmap-with-taco-original}.
\section{Related Work} \label{related_works}

Code generation for tensor algebra has been extensively researched.
This area of research can be primarily divided into two subareas --- sparse and dense.
First, we discuss the related work on code generation and optimization techniques for sparse tensor algebra and then move to the ones for dense. 

\subsection{Sparse Tensor Algebra}

Automated sparse code generation~\cite{aartbik:93,compiler_in_mlir,kjolstad:2017:taco,aartbik93_data_structure_selection} is a heavily-researched topic. 
Even though these methods are highly effective, they lack fine-grained optimizations like ours that applies across kernels to reduce the time-complexity of the computation.
Ahrens \etal~\cite{peterahrens} proposed splitting large tensor expressions into smaller kernels to minimize the time-complexity.
The Sparse Polyhedral Framework~\cite{LaMielle2010EnablingCG,sparse_polyhedral_framework,sparse_polyhedral_framework2} employs an inspector-executor strategy to transform the data layout and schedule of sparse computations to achieve locality and parallelism.
Kurt \etal~\cite{tiled_sparse} improved SpMM and SDDMM kernels by optimizing their tile sizes using a sparsity signature. 
However, these methods do not consider loop nest restructuring transformations to improve data locality across kernels.

Athena~\cite{athena}, Sparta~\cite{sparta}, and HiParTi~\cite{hiparti} are techniques which provide highly optimized kernels for sparse tensor operations and contraction sequences that shows significant performance improvements. 
Kernel fusion has been used in FusedMM~\cite{fusedMM} to accelerate SDDMM and SpMM used in graph neural network applications. Their transformation is structurally analogous to \system, but is specific to graph embeddings, and further performs kernel-specific optimizations.
None of these prior techniques handle arbitrary sparse tensor expressions supporting a variety of input formats.

\subsection{Dense Tensor Algebra}
Optimizations for computations over dense tensors have been well-studied for decades.
Numerous loop optimizations for dense tensor contractions~\cite{Cociorva,saday1,saday2005,saday2006,Krishnan2003DataLO,saday2001,saday2002} for CPUs and tensor contractions for GPUs~\cite{7349652,ABDELFATTAH2016108} have been proposed that exhibits superior performance. 
However, these transformations are not directly applicable to sparse tensor algebra since there data access restrictions for sparse tensors and the non-affine nature of loop nests.






\section{Conclusion} \label{discussion}

We presented \system, a loop restructuring framework for sparse tensor algebra programs. 
\system improves the performance of sparse computations by reducing time complexity and enhancing data locality.
\system enables kernel distribution and loop fusion and achieves significant performance improvements for real-world benchmarks.
The new scheduling transformations introduced by \system expands the scheduling space of sparse tensor applications and facilitates fine-grained tuning. 


\begin{acks}                            
  We appreciate the feedback from the anonymous reviewers for their suggestions and comments that helped to improve this paper.
  We would also like to thank Fredrik Kjolstad for the valuable discussions we had regarding the \system transformation.
  This work was supported in part by the \grantsponsor{nsf}{National Science Foundation}{} awards \grantnum{nsf}{CCF-1908504} and \grantnum{nsf}{CCF-1919197}.
  Any opinions, findings, and conclusions or recommendations expressed in this paper are those of the authors and do not necessarily reflect the views of the \grantsponsor{nsf}{National Science Foundation}{}.
\end{acks}

\balance
\bibliography{bibfile}

%

\end{document}